%% file: paper3.tex
\documentclass[useAMS,usenatbib]{mnras}
\usepackage{rotating}

\bibpunct{(}{)}{;}{a}{}{,}

\def\yr{{\,{\rm yr}}}
\def\ang{{\rm \AA}}
\def\imfs{$\alpha_3$}
\def\slug{{{\sc{slug}}}}
\def\sb{{{\sc{starburst99}}}}
\def\msun{{\rm{M}_{\sun}}}

\title[IMF diagnostics from UV spectroscopy]
{Theoretical predictions for IMF diagnostics in UV spectroscopy of star clusters }
\author[G. Ashworth, M. Fumagalli, et. al.]
{G. Ashworth $^{1,2}$, Michele Fumagalli $^{1,2}$, Angela Adamo $^{3}$, Mark R. Krumholz $^4$\\
$^1$Institute for Computational Cosmology, Durham University, South Road,
Durham DH1 3LE, UK\\
$^2$Centre for Extragalactic Astronomy, Durham University, South Road,
Durham DH1 3LE, UK\\
$^3$Department of Astronomy, The Oskar Klein Centre, AlbaNova University Centre, Stockholm University, SE-106 91 Stockholm, Sweden\\ 
$^4$Research School of Astronomy \& Astrophysics, Australian National University, Canberra, ACT 2611, Australia\\}
\pagerange{\pageref{firstpage}--\pageref{lastpage}} \pubyear{}

\pubyear{2018}

\begin{document}

\topmargin = -0.5cm

\maketitle

\label{firstpage}

\begin{abstract}

We explore the possibility of using UV spectroscopy in combination with broad-band photometry as diagnostic tools for understanding the shape of the Initial Mass Function (IMF) in unresolved stellar populations. Building on our previous work, we extend the Stochastically Lighting Up Galaxies code (\slug) to include a high-resolution UV spectral synthesiser and equivalent width calculation capabilities.
We first gain a qualitative understanding of how UV spectral features behave as the parameters that define a star cluster in \slug\ (mass, age, extinction, and IMF slope \imfs) are changed. 
We then exploit Bayesian inference techniques to recover the \imfs\ values for clusters simulated with \slug, using mock observations of these clusters comprised of broad-band photometry and equivalent width measurements of a selection of UV spectral features. We find some improvement when compared to attempts using broad-band photometry alone (with the interquartile range of the \imfs\ posterior PDF shrinking by $\simeq32\%$), although we still do not yet fully break the known degeneracy between the cluster mass and \imfs.
Finally, we make predictions about how effective real observations will be by quantifying our ability to constrain \imfs\ as a function of limiting equivalent width. We find that observations sensitive to a modest equivalent width of $\simeq9\,\ang$ are sufficient to improve the recovery of the IMF slope parameter by $\simeq32\%$ (interquartile range of posterior PDF median residuals), moving to $\simeq39\%$ when we include all the significant spectral features in the wavelength range $900-3000\,\ang$.

\end{abstract}


\section{Introduction} \label{sec:introduction}

Young, bright, massive stars are responsible for much of the ultraviolet (UV) flux of a stellar population, and it is from these stars that many distinctive features in the UV region of the spectrum originate. The number of these stars that exist in a cluster is dependent on the upper end slope of the initial mass function (IMF) \citep[e.g.][]{Salpeter1955-IMF}. The IMF describes the distribution of stars by mass in a stellar population, and it is a core ingredient in our understanding of many astrophysical phenomena. As these massive stars contribute strongly to the UV region of the spectrum, this region holds many clues which can aid investigations into the shape of the IMF. Accurately modelling the UV region of a star cluster's spectrum is therefore important if we are to improve our understanding of the IMF from observations of unresolved stellar populations. It is such a task that we undertake over the course of this paper.

Describing the shape of the IMF is an important goal as it is the IMF that defines the makeup of stellar populations. Its shape is generally considered to be universal \citep{Bastian2010-IMF}, with most observations suggesting that, at its simplest, it takes the form of a power law of slope $\sim -2.3$ \citep{Salpeter1955-IMF}, assuming a more shallow slope at low stellar masses \citep{Kroupa2001-IMF,Chabrier2003-IMF,Massey1995-IMF,Bastian2010-IMF}.

However, the concept of a universal IMF is not without challenge. There exist many studies which imply variation in the shape of the IMF, and many of these concentrate on the upper end slope. These studies range from theoretical arguments based on temperature and metallicity \citep{Adams1996-IMFTheory,Bonnell2006-IMFJeansMass,Krumholz2011-FragIMF} through to the Integrated Galactic Initial Mass Function (IGIMF) \citep{Kroupa2001-IMF,Kroupa2003-IGIMF,Weidner2011-IGIMF,Kroupa2013-StellarSubStellar} which is based on the assumption that the most massive star that can form in a star cluster is related to the mass of that cluster (i.e. the upper limit of the IMF). In a recent study, \cite{Dib2017-VariableIMF} find suggestions of a variable IMF in the Milky Way through `lonely' O star fractions, and \cite{Schneider2018-BigStars} infer a shallow high-mass end IMF slope for 30 Doradus. However, no evidence of IMF variation has been widely accepted in the literature, and a variety of alternative explanations have been proposed to many of these variable IMF indications, from the application of stochastic sampling of the IMF \citep{Fumagalli2011-SLUGLetter} to the reported observations of isolated massive OB stars \citep{Oey2013-SoloOBStars,Lamb2016-RIOTS}.

In this paper, we continue the work started in \cite{Ashworth2017-SLUGandLEGUS}. In that study, we investigated the possibility of constraining the IMF from broad-band photometry using the Stochastically Light Up Galaxies \citep[\slug;][]{daSilva2012-SLUGi,daSilva2014-SLUGii,Krumholz2015-SLUGiii,Ashworth2017-SLUGandLEGUS} stellar population synthesis (SPS) code. Using {\it Hubble Space Telescope} photometry from the Legacy ExtraGalactic UV Survey \citep[LEGUS;][]{Calzetti2015-LEGUSi}, combined with Bayesian analysis exploiting a large library of \slug\ models, we demonstrated that accurate recovery of the IMF was non-trivial. Using mock clusters generated with \slug, we went on to quantify the degeneracy between the mass and the IMF, and showed that using broad-band photometry alone, without accurate knowledge of the cluster mass, we could not put tight constraints on the slope of the IMF.

In this work, we include the UV region of the spectrum in addition to the broad-band photometry used previously. As UV features arise in the spectra of young, hot, and massive stars, including this region of the spectrum in the analysis should allow us to constrain the IMF's high-end slope more closely than was shown to be possible using broad-band photometry alone. Other studies have made use of this fact, for example, in the recent work of \cite{Wofford2014-MassiveStars}, the authors use UV spectroscopy to constrain the high-mass end of the IMF (finding it not to be shallower than the Salpeter value). Previous work that has combined SPS modelling with UV spectral features to determine the IMF includes, e.g., \citet{Pettini2000-UVSpec}, \citet{Steidel2004-Survey}, and \citet{Quider2009-UVSpec}, although these studies tend to perform more traditional statistical analysis. For instance, comparing deterministic SPS model spectra by eye to galaxy spectra calculated from discrete sets of physical parameters.

As in our previous work, we make use of \slug\ to perform this study. The paper is structured in the following way.
In Section~\ref{sec:hires}, we describe new extensions made to the \slug\ code in order to produce high-resolution UV spectra. These modifications are then used to construct several libraries of simulated clusters with a wide coverage of the mass-age-extinction-IMF parameter space. 
In Section~\ref{sec:models}, we continue with a qualitative exploration how the behaviour of the physical parameters that define the stellar populations maps onto UV spectral features and broad-band photometry.
Next, in Section~\ref{sec:mockob}, we produce photometry and equivalent width measurements for a selection of simulated clusters that form a set of mock observations. These observations are then analysed using the Bayesian inference methods used in previous studies \citep{Krumholz2015-SLUGandLEGUS,Ashworth2017-SLUGandLEGUS}, which we now extend to include spectral features as additional input data. 
Finally, we make predictions as to what spectral features are of importance when planning observations, and whether observations would be able to recover the high-mass IMF slope of an unresolved stellar population using the methods presented in this paper.

\input{table_models.tex}
\input{table_lines.tex}

\section{Implementing high-resolution UV spectra in \slug}\label{sec:hires}
To generate our mock star clusters, we use the \slug\ code \citep{daSilva2012-SLUGi,daSilva2014-SLUGii,Krumholz2015-SLUGiii,Ashworth2017-SLUGandLEGUS}. \slug\ is a stochastic stellar population synthesis (SPS) code that allows the user to generate star clusters and galaxies by randomly drawing stars from an input IMF, as well as drawing star clusters themselves from a cluster mass function when populating galaxies. \slug\ also has the capability to randomly sample the star formation history when populating galaxies.
As part of the SLUG suite of software\footnote{Available at \url{www.slugsps.com}}, we distribute a Bayesian analysis toolset by the name of {\sc bayesphot}. The {\sc bayesphot} tool uses Bayesian inference techniques to calculate posterior probability distribution functions (PDFs) of physical parameters that define the cluster from a set of input observables \citep[generally broad-band photometry;][]{Krumholz2015-SLUGiii}. This is accomplished through the use of kernel density estimation (KDE) techniques, combined with a large library (which we refer to as the `training set') of \slug\ models.

\subsection{Stellar atmosphere grids}

Up until now, \slug\ has been capable of producing spectra with a resolution of $\sim 10\,\ang$. However, this is not a high enough resolution to allow us to make use of the equivalent widths of absorption line features in our analysis. As it is these features that we wish to use to infer the slope of the high-mass end of the IMF, we have implemented a high resolution spectral capability in the \slug\ code. Along with this new capability, the associated machinery required to calculate equivalent widths for arbitrary line features and use them in Bayesian inference has also been implemented.

The highest spectral resolution that \slug\ is capable of producing is limited to the resolution of the stellar atmospheres to which it has access. Therefore, in order to produce high resolution spectra, \slug\ must have access to high resolution stellar atmosphere models in the wavelength region of interest. To this end, we modified \slug\ to handle the high-resolution ($\sim 0.4\,\ang$) atmospheres of \cite{Leitherer2010-SB99Spectra}, which are distributed as part of \sb\ \citep{Leitherer1999-STARBURST99}. These atmospheres cover a wavelength range of $900-3000\,\ang$. 

This new spectral synthesiser option, dubbed {\tt SB99HRUV} in \slug, covers the UV region of the spectrum ($900-3000\,\ang$) using the following atmospheres:

\begin{itemize}
\item Stars classified as Wolf-Rayet stars use the Potsdam PoWR atmospheres \citep{Grafener2002-PoWR,Hamann2004-PoWR,Sander2012-PoWR}, specifically the revision shipped with \sb.
\item We use Kurucz \citep{Kurucz1992-Atmo,LeJeune1997-Kurucz} atmospheres for all stars with $M < 5\,\msun$, for stars with $T < 170000$ K in the mass range $5-10\,\msun$, and for stars with $T < 22000$ K at masses $>10\,\msun$. 
\item For all other stars, we use the high resolution IFA atmospheres \citep{Leitherer2010-SB99Spectra}.
\item Stars that fall outside of these atmosphere grids are treated as black bodies, although this is a rare occurrence given our choice of tracks.
\end{itemize}

For the rest of the wavelength range of the spectrum, we use the standard lower resolution atmospheres (equivalent to selecting the default \texttt{SB99} spectral synthesizer option, which mimics the atmosphere selection of \sb) that are employed by \slug. Further information is available in \cite{Krumholz2015-SLUGiii} and the \slug\ user manual. The core differences between this and the high resolution region described above are the use of the Hillier atmospheres \citep{Hillier1998-Atmo} for Wolf-Rayet stars, and the use of Pauldrach OB atmospheres \citep{Pauldrach1998-Atmo} instead of the new IFA atmospheres.

All the atmospheres used during the work performed in this paper are those that are distributed with either \slug, or in the case of the PoWR atmospheres and the high resolution IFA UV atmospheres, those that are distributed with \sb. Finally, we stitch the high and low resolution spectra together, replacing the $900-3000\,\ang$ region of the low resolution spectrum with the high resolution spectrum calculated using the atmospheres listed above.

In our calculation we assume the Padova stellar tracks including thermally pulsating AGB stars \citep{Vassiladis1993-PadovaAGB,Girardi2000-Padova,Vazquez2005-Padova}. We must however consider the effects of the limits of these tracks on the IMF we have selected. Indeed, we show in Section~\ref{sec:modelfeatures} how there can be significant differences between clusters generated with a $30\,\msun$ mass cutoff compared to an IMF which cuts off at $120\,\msun$, for example. There are, however, some observational indications for stars with masses greater than the assumed cutoff \citep[for example, the recent work of][who find evidence for stars with masses up to $200\,\msun$ in 30 Doradus]{Schneider2018-BigStars}
which warrant further exploration before applying our formalism to real data.

\subsection{Selecting atmospheres}

When selecting the high-resolution IFA atmospheres, we have implemented both the nearest-neighbour algorithm used by \sb, and a new interpolation scheme whereby the atmosphere for a particular star is calculated through a barycentric interpolation. This can be toggled by simply changing a Boolean value in the high resolution UV spectral synthesiser before compiling \slug. The nearest-neighbour scheme allows direct comparison with \sb, and is also less computationally intensive. 

There are several steps to our new interpolation scheme. We begin with the parameters of the star for which we wish to obtain an atmosphere. First we find the two nearest neighbour atmospheres in the temperature-surface gravity plane that exist within the atmosphere library. We then iterate through the next nearest neighbours until we satisfy the conditions required to enclose the star within a triangle of three atmospheres within this plane. Having formed this triangle of nearest neighbour atmosphere models, we find the barycentric coordinates of the star of interest within this triangle, and then use these to weight the atmospheres on each vertex of the triangle. The weighted combination of these atmospheres gives us an interpolated model for the star. If the triangle cannot be closed, we default to the nearest-neighbour method. We use this interpolation scheme during this paper rather than the nearest neighbour approach due to the sometimes large gaps in the parameter space. The definition of `nearest neighbour' is the same one used by \sb\ for these atmospheres, with each direction (in the surface gravity - temperature plane) weighted differently \citep[with the surface gravity axis weighted by a factor of 5;][]{Leitherer2010-SB99Spectra}.

To test the scheme, a test star was selected, and then removed from the atmosphere grid. The spectrum of the nearest neighbour star and the interpolated atmosphere were then compared to the true atmosphere of the missing star. The interpolated star lay noticeably closer to the true value across most absorption features
\footnote{Even when using the basic nearest-neighbour atmosphere selection scheme, results produced by \slug\ will not be directly comparable to \sb\ due to the differing track methods that \slug\ employs. However, using an older revision of the \slug\ code, and using the nearest-neighbour atmosphere selection method, we were able to validate our implementation by reproducing the rectified and non-rectified spectra produced by \sb\ for a selection of test clusters.}.
Our work is based on mock observations constructed with a consistent choice of interpolation scheme. These mock observations are then analysed using libraries of \slug\ models constructed using the same interpolation scheme. Thus, we defer the study of which scheme performs best when applied to observations to future work.

We further note that, due to the low density of points in some regions of the temperature-surface gravity plane, certain combinations of parameters can lead to significant differences in flux between the high resolution spectrum and the low resolution spectrum that we stitch into, as in \sb. This leaves discontinuities at the borders of the high resolution region in some cluster parameter configurations, and affects photometry for filters that cross into this region. To ensure smooth transitions, we integrate a region in both the low resolution and high resolution spectra (covering $2400-2900\,\ang$) and scale the high resolution spectrum by the ratio between these integrals, so that we can eliminate any large discontinuities at the points where the two spectra join. It is recommended, however,  that users planning to compare \slug\ photometry alone to real data use the low resolution spectrum.

\subsection{Equivalent width calculations}

As the high-resolution IFA spectra are provided in both line and continuum form, we are able to normalise the spectrum without the difficulties of attempting to create a pseudo-continuum around each line feature. Therefore, we use only the rectified spectrum (as is output by \sb) to calculate equivalent widths, which gives us a normalised spectrum consisting of features belonging only to the stars covered by the new high resolution IFA atmospheres (the massive O and B type stars). This ensures that a true continuum is adopted in the equivalent width calculations rather than an empirically-determined pseudo-continuum. As O and B stars dominate the signal for the features analysed in this work, this treatment provides an excellent approximation overall. In simple testing using an approximate normalisation of the spectra for the other stellar types as performed by \sb, we find that the majority of our ``independent'' line list (see Table~\ref{tab:lines}) are not strongly affected. The strongest effects are apparent around the ages where Wolf-Rayet stars contribute to the spectrum. Even then, only our Si{\sc i} and O{\sc vi}b features are strongly affected (with the latter present in the representative line list and the former in the independent line list). Between $10^5\,\yr$ and $10^7\,\yr$ we see only minor differences for other lines in our independent line list (less than ten percent when moving to the oldest clusters). The exception is for Mg{\sc i} and Mg{\sc ii} where we see differences of approximately $20-30\%$. There will also be an effect on any photometric filters which overlap the high resolution region. However, we must note that these differences will change if newer versions of the Wolf-Rayet atmospheres are used.
Overall, the use of O and B type stars alone is an acceptable approximation for this theoretical study, and the inclusion of more information in the spectrum in the form of additional stellar types may in fact assist in the recovery of the cluster physical parameters.
It should also be noted that these rectified spectra are not extincted, and contain no nebular emission. We do, however, run \slug\ with nebular emission calculations enabled for the work in this paper, and both it and extinction are applied to the photometric filters and the raw non-rectified spectrum. 
Overall, we must stress that, even with the limitations detailed above, as we are comparing \slug\ mock observations to libraries of other \slug\ models, this approach is internally consistent for this study.

Finally, the equivalent width is calculated by interpolating linearly between each point in the spectrum (as they are the bin-centres of variable width bins, this is a best approximation), and then integrating under the resulting function across the integration region for a given spectral feature.
Further information on the use of these new features is available in the \slug\ documentation\footnote{\url{http://slug2.readthedocs.io}}. The code is currently publicly available, including the high-resolution IFA atmospheres. The PoWR Wolf-Rayet atmospheres will be added to the distribution when new versions from the PoWR group are made available (W.~R. Hamann 2017, Private Communication) as these supersede the versions provided with \sb, which are used here.

\begin{figure*}
\begin{center}
\begin{tabular}{cc}
\includegraphics[scale=0.6,trim={0.45cm 0 0.45cm 0},clip]{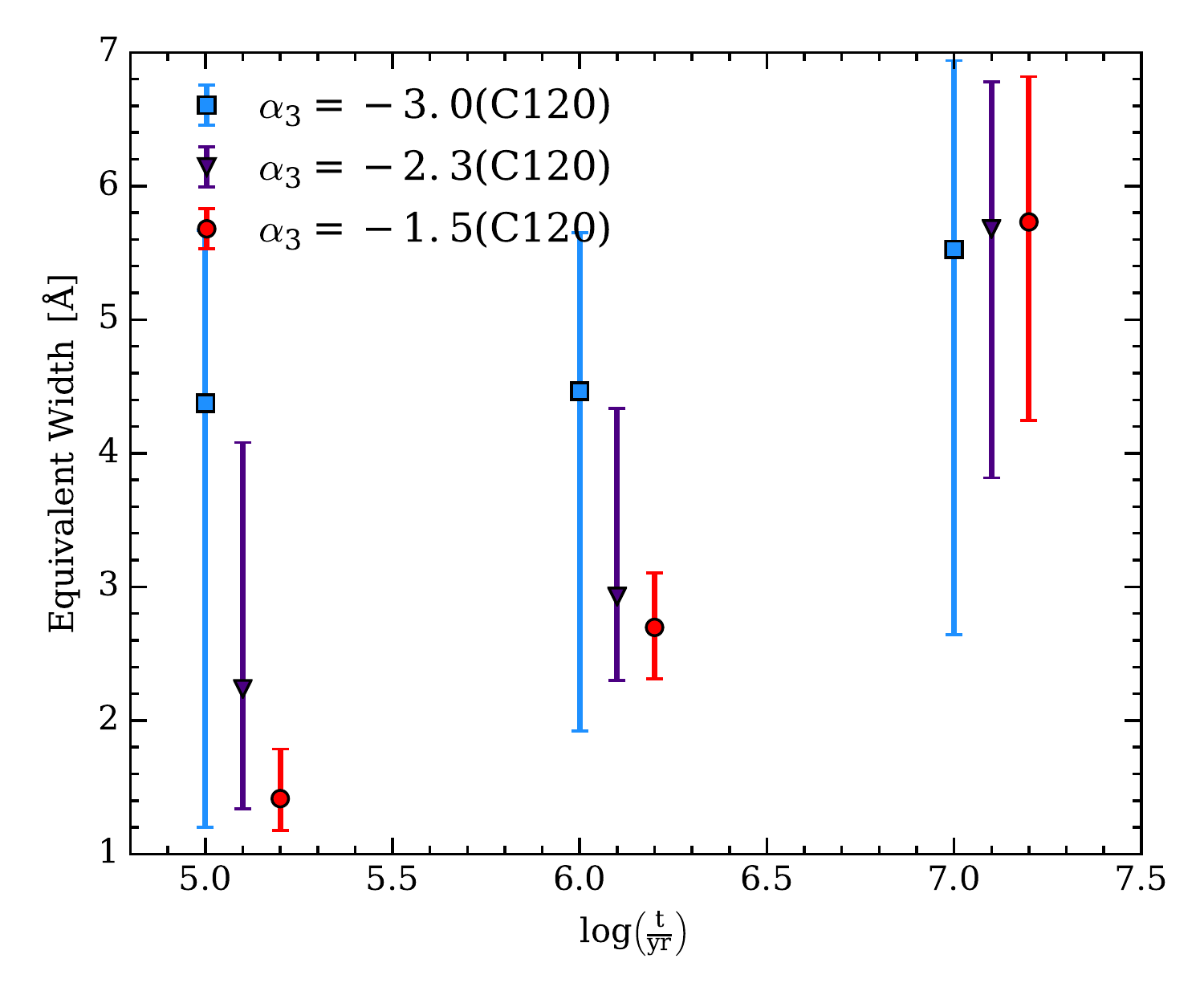} & \includegraphics[scale=0.6,trim={0.45cm 0 0.45cm 0},clip]{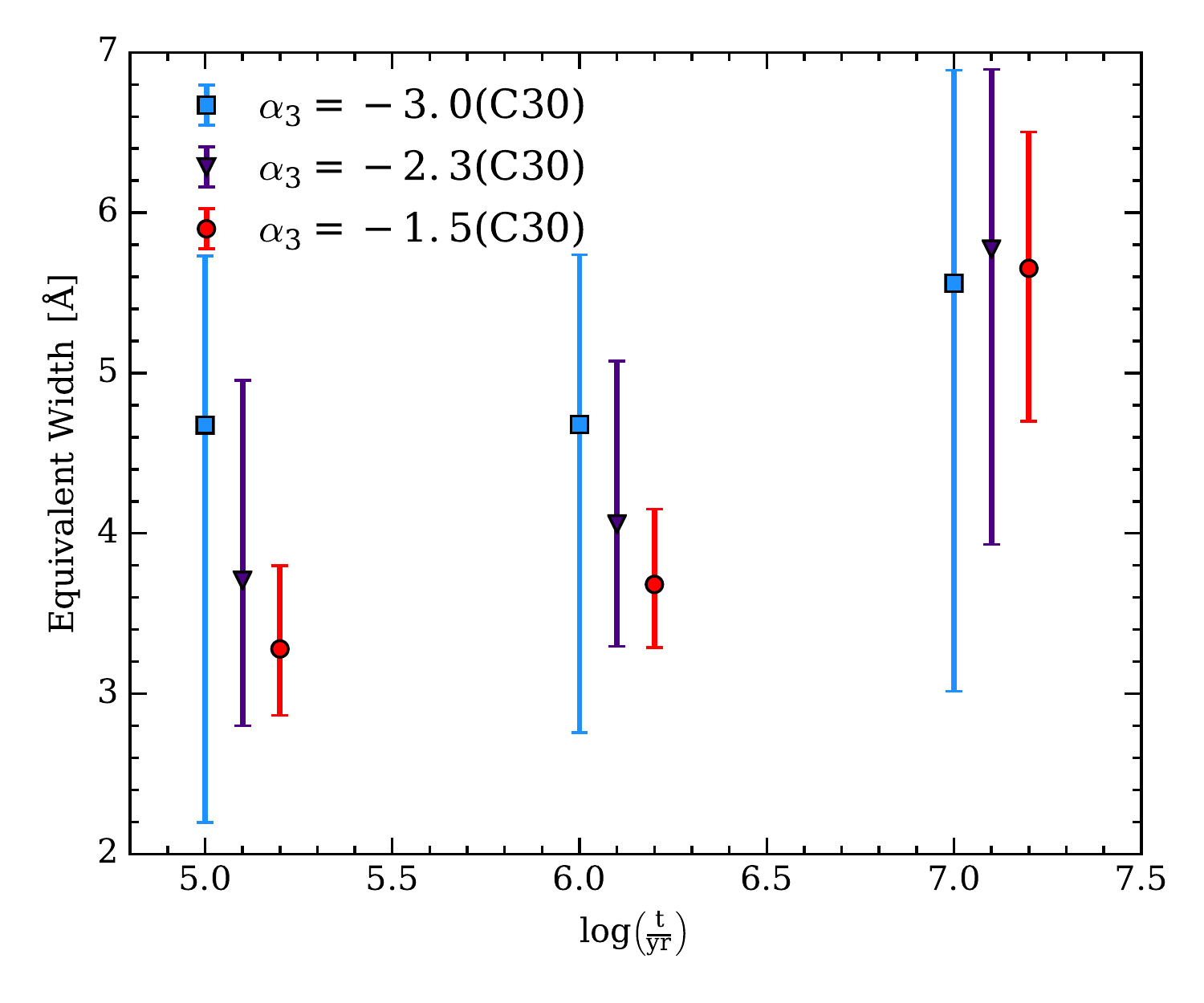} \\
(a) N{\sc iv}: $\log\left(\frac{M}{\msun}\right)=3$, IMF cutoff at $120\,\msun$ & (b) N{\sc iv}: $\log\left(\frac{M}{\msun}\right)=3$, IMF cutoff at $30\,\msun$   \\[6pt]
\includegraphics[scale=0.6,trim={0.45cm 0 0.45cm 0},clip]{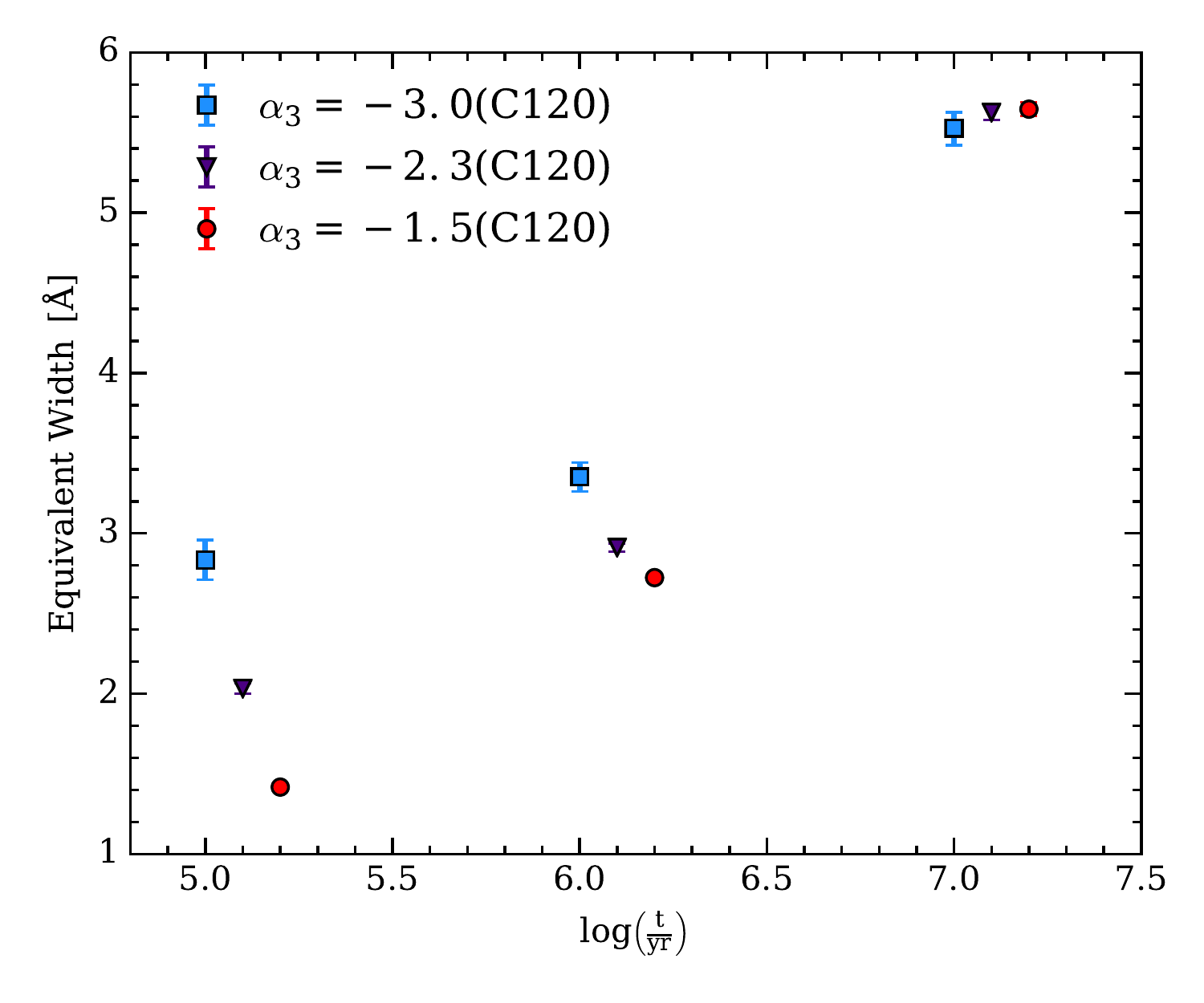} & \includegraphics[scale=0.6,trim={0.45cm 0 0.45cm 0},clip]{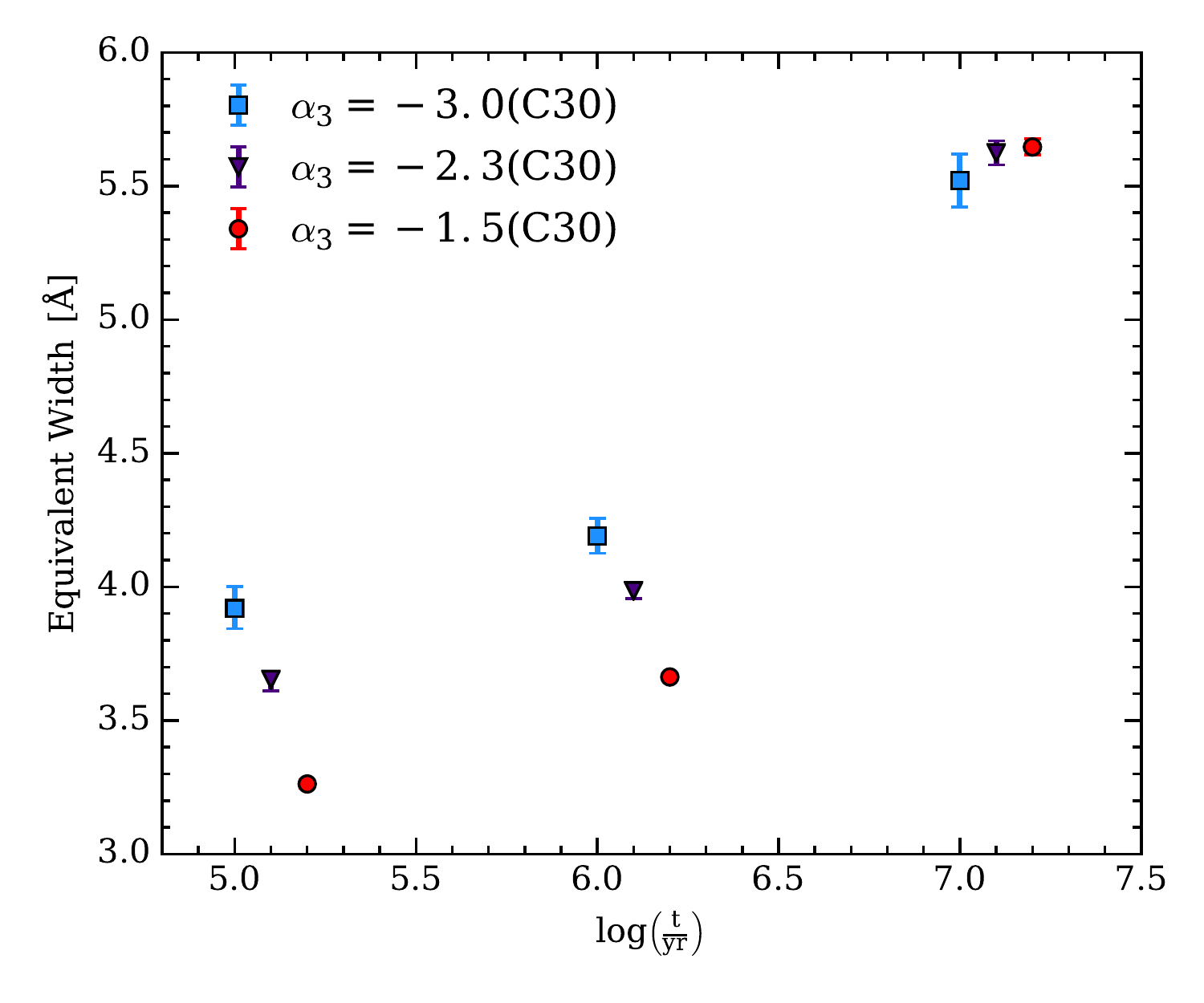}\\
(c) N{\sc iv}: $\log\left(\frac{M}{\msun}\right)=6$, IMF cutoff at $120\,\msun$    & (d) N{\sc iv}: $\log\left(\frac{M}{\msun}\right)=6$, IMF cutoff at $30\,\msun$  \\ [6pt]
\end{tabular}
\end{center}
\centering
\caption{Evolution of the equivalent width of N{\sc iv} with cluster age for a range of IMF upper-end slopes. Panels (a) and (b) are clusters with a mass of $10^3\,\msun$, whereas panels (c) and (d) are clusters with masses of $10^6\,\msun$. Panels (a) and (c) have the IMF cutoff at $120\,\msun$, whereas panels (b) and (d) have the IMF cutoff at $30\,\msun$. The points correspond to the median value from the set of realisations, and the error bars cover the fifth to ninety-fifth percentile range. Blue squares correspond to $\alpha_3=-3.0$, purple triangles correspond to $\alpha_3=-2.3$, and red circles correspond to $\alpha_3=-1.5$. The leftmost point of each trio is positioned at the correct age value, and the others are shifted slightly to aid readability.}
\label{fig:singlefeature}
\end{figure*}

\begin{figure*}
\begin{center}
\begin{tabular}{cc}
\includegraphics[scale=0.6,trim={0.45cm 0 0.45cm 0},clip]{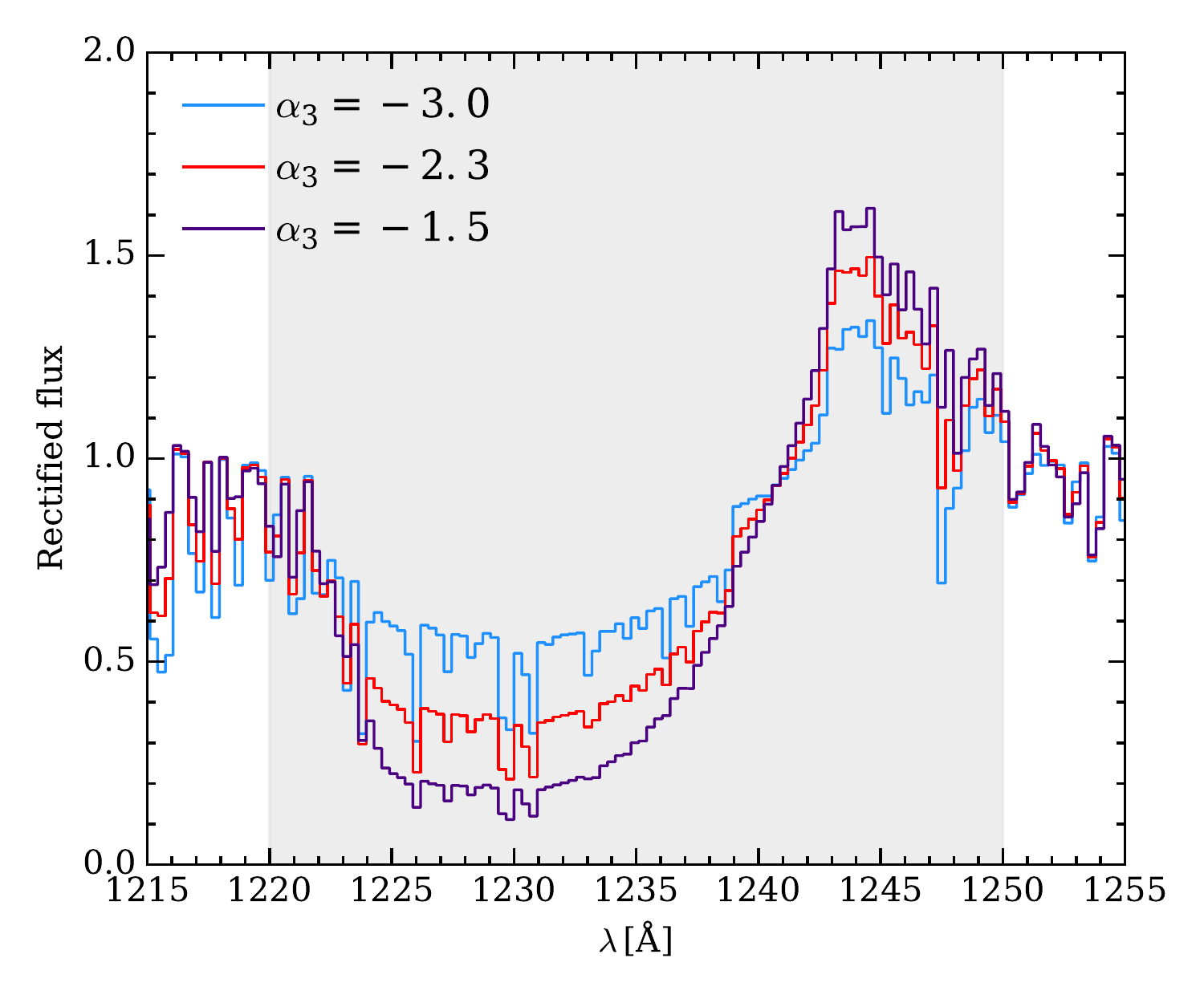} & \includegraphics[scale=0.6,trim={0.45cm 0 0.45cm 0},clip]{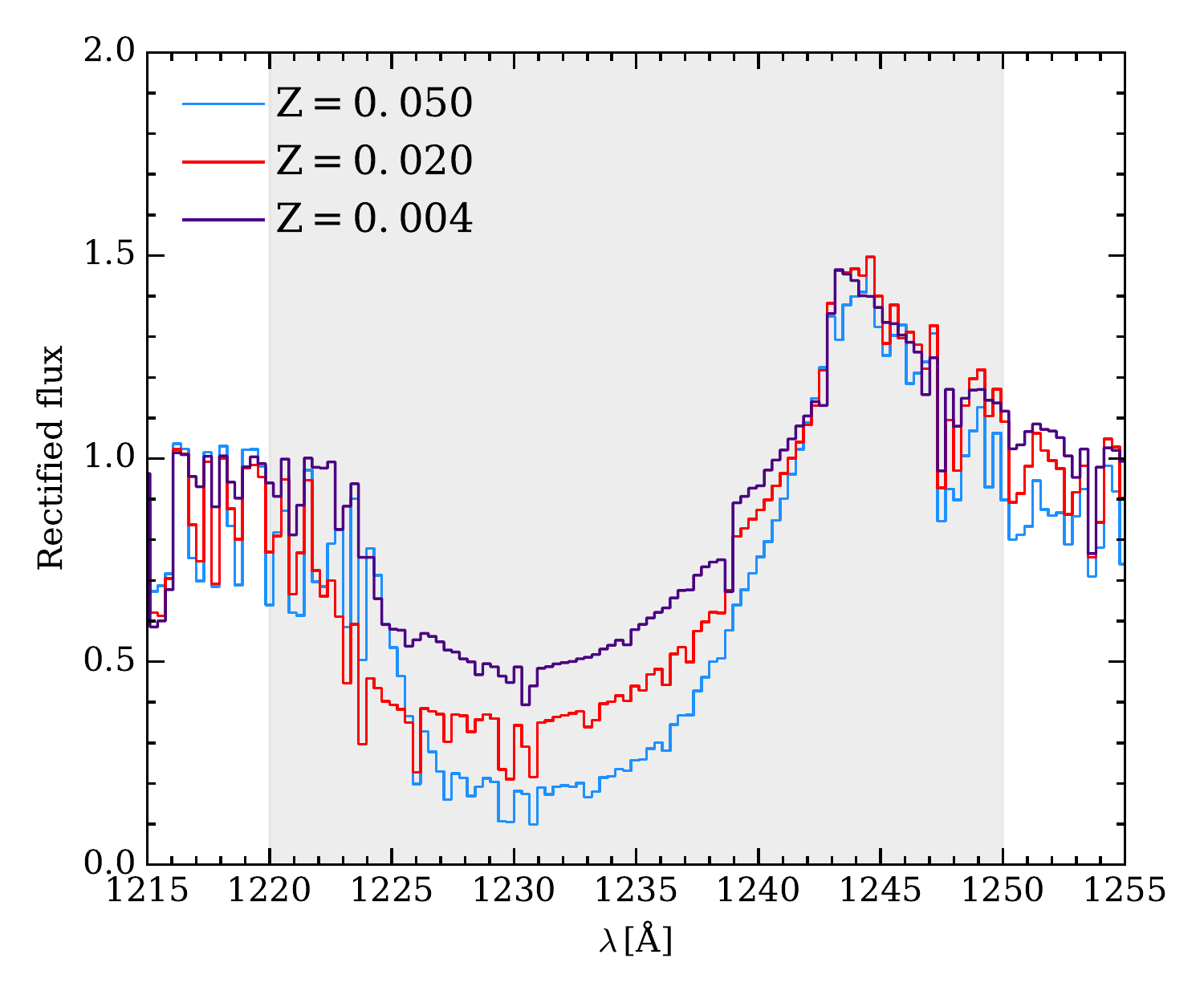} \\
(a) N{\sc v} - $\Delta\alpha_3$ & (b)   N{\sc v} - $\Delta Z$ \\[6pt]
\end{tabular}
\end{center}
\centering
\caption{Renormalised stacked spectra of 1000 model star clusters of mass $10^6\,\msun$ at an age of $10^5\yr$. In panel (a), each set of 1000 was generated at a different IMF ($\alpha_3=-3.0,-2.3,-1.5$), at solar metallicity. In panel (b), each set of 1000 was generated using a different metallicity, with $Z=0.004,0.020,0.050$. All models were generated using the canonical Kroupa IMF as a base IMF shape. The spectrum shows the region around the N{\sc v} line, with the integration region used for the calculation of the equivalent width outlined in a light grey colour. We see the line increase in strength as the IMF becomes shallower. Note also the degeneracy in the absorption region between the IMF and the metallicity. The emission region is unaffected by changing metallicity.}
\label{fig:spectra_dz}
\end{figure*}

\begin{figure*}
\begin{center}
\begin{tabular}{cc}
\includegraphics[scale=0.6,trim={0.45cm 0 0.45cm 0},clip]{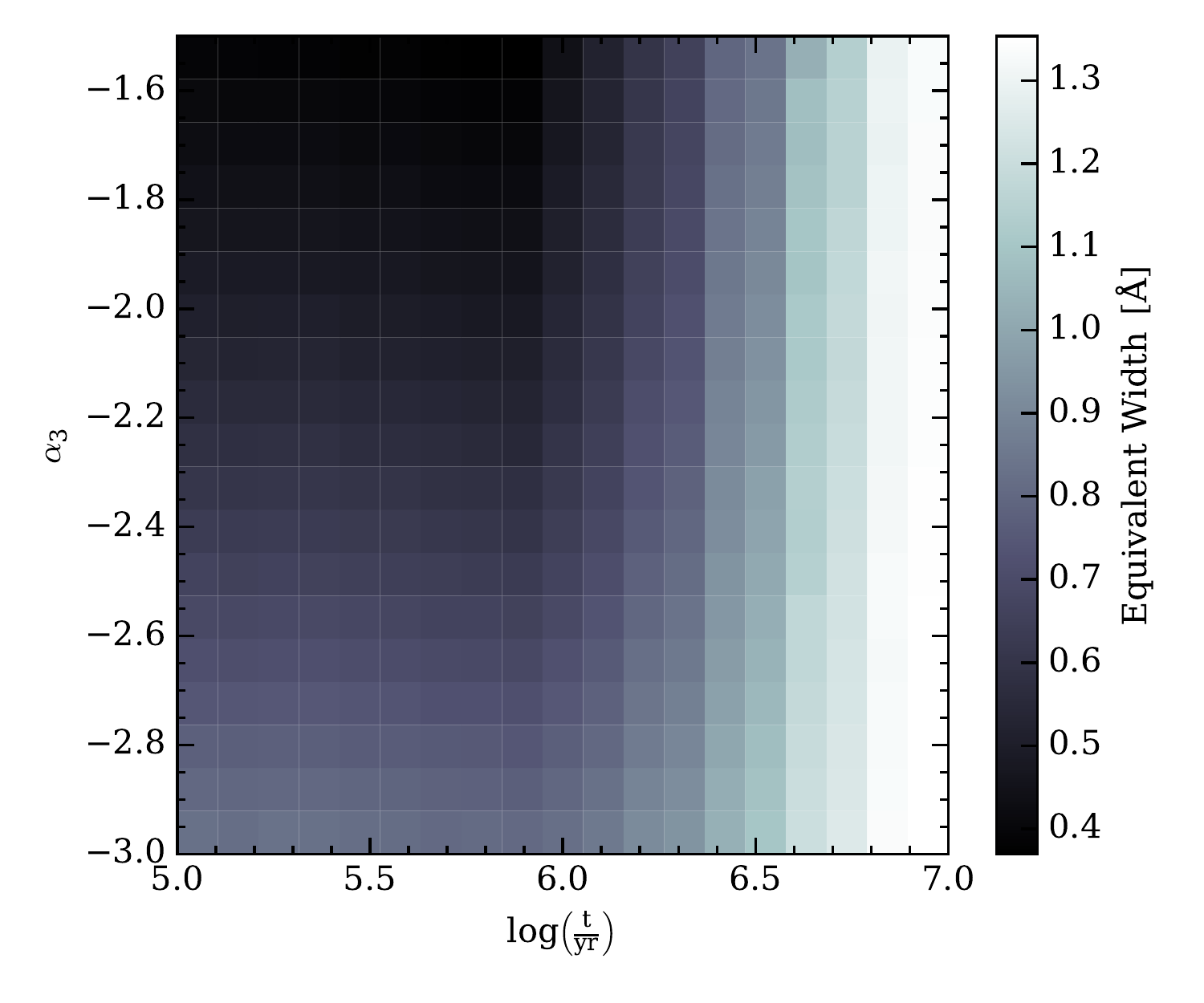} & \includegraphics[scale=0.6,trim={0.45cm 0 0.45cm 0},clip]{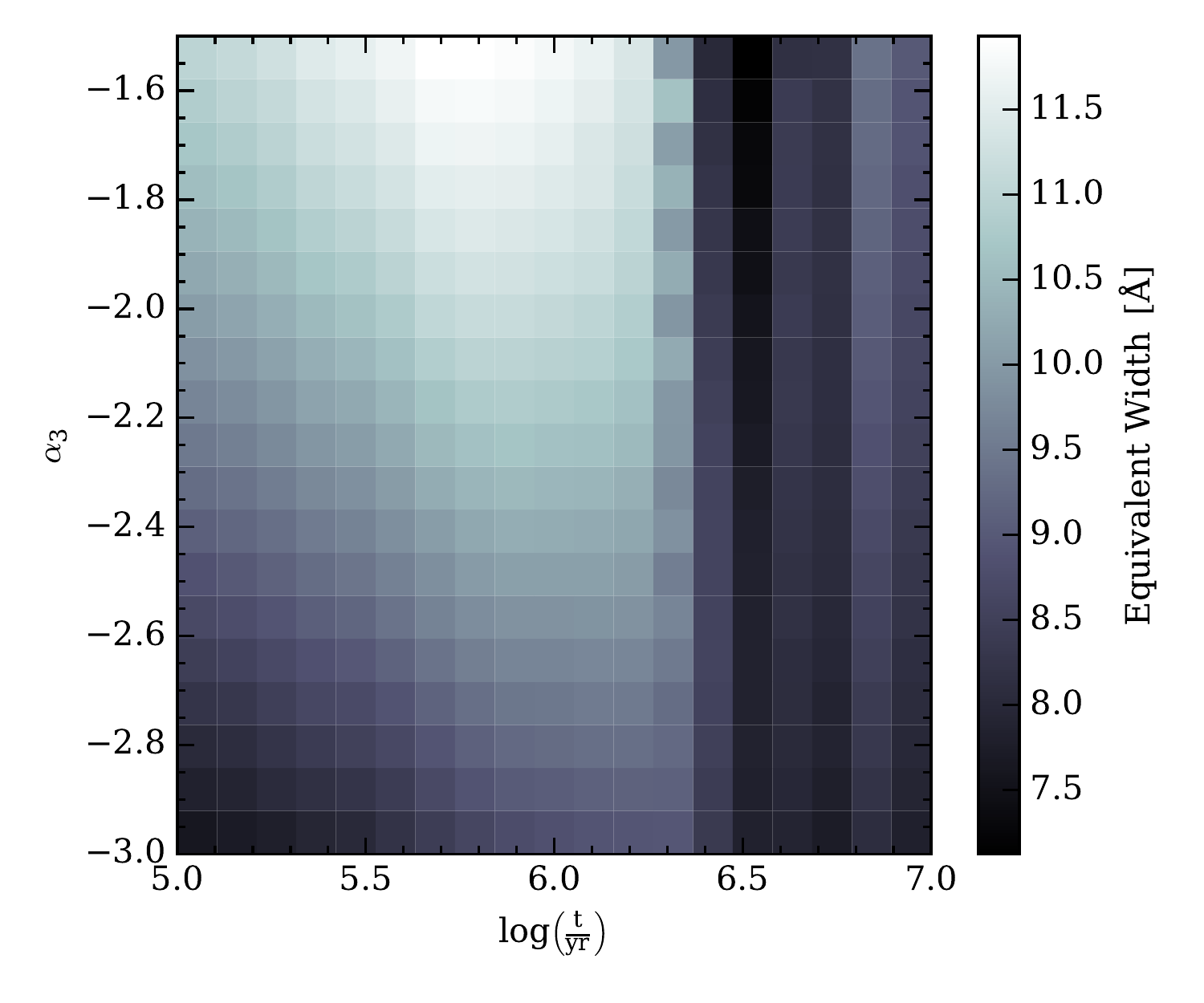} \\
(a) C{\sc iii} & (b)   C{\sc iv}a  \\[6pt]
\end{tabular}
\end{center}
\centering
\caption{Two dimensional plots of equivalent width versus changes in cluster age or IMF slope for two spectral regions (C{\sc iii} and C{\sc iv}a), where the equivalent width reported for a grid cell is the median of the equivalent widths for the clusters that reside in that area of parameter space. In both panels we see that the equivalent width of these lines may be used to differentiate the various possible values of \imfs\ at young ages. In addition, the combination of the two lines provides us with a way to also differentiate the age of the cluster in conjunction with the value of \imfs.}
\label{fig:squares}
\end{figure*}

\section{Grids of high-resolution \slug\ models}\label{sec:models}

In order to carry out the work described in this paper, we must generate a large selection of \slug\ models. These models will be used for empirical study, to form mock observations, and to form the large library of simulations required by {\sc bayesphot}. The parameters that define these different models are detailed in Table~\ref{tab:modelparams}.

The `Library' is the large set of models that makes up the `training set' we use for our Bayesian analysis.
The Library contains $5\times 10^7$ model star clusters, with their masses drawn from a flat distribution between $5\times 10^2\,\msun$ and $5\times 10^5\,\msun$, and their ages drawn from a flat distribution between $5\times 10^4\,{\rm yr}$ and $5 \times 10^7\,{\rm yr}$. The extinction, $A_{\rm V}$ is drawn from a flat distribution between $0$ and $3$, and the IMF high-mass slope \imfs\ is drawn from a flat distribution between $-3.2$ and $-1.3$. The actual shape of the distribution is not important during the Bayesian analysis, but a flat distribution makes sense in this context as the clusters we use to generate our input photometry are also generated using flat distributions. Nebular emission is calculated using an ionising photon fraction $\phi=0.73$, the default in \slug. It must be noted once again, however, that nebular emission (much like extinction) is not included in the rectified spectrum used for calculating equivalent widths from the combinations of high resolution stellar atmospheres. This does mean that any nebular emission features are absent from the spectra we use to calculate the equivalent widths. However, much of the increased flux from the nebular emission, and the effects of extinction, would be negated by the normalisation process. In addition, as the observations are generated with the same restrictions as the models used in the Bayesian analysis the work is self consistent.

The `Mock Obs.' models are those whose photometry and equivalent widths form our set of mock observations that we use in our Bayesian analysis.
Their masses are drawn from a slightly more narrow set of distributions in mass, age, and IMF than the library, as shown in Table~\ref{tab:modelparams}. This means we have no mock clusters situated right at the edge of the library in any part of parameter space (save for extinction, where we cannot go negative), or even outside of it. This is a desirable property when studying clusters with {\sc bayesphot}.
`Sets' 1 to 6 are models used for studies of both individual lines and the behaviour of the equivalent width for varying age, mass, \imfs, and metallicity. Each set is comprised of 1000 realisations, enough for the median equivalent widths to be acceptably converged.
Finally, the `squares' set is a group of models intended to be used to investigate the changes in equivalent width in the age-\imfs\ plane. These models are drawn from flat distributions in $\log t$ and \imfs.

All of these sets of models have both photometry, along with equivalent widths for the features listed in Table~\ref{tab:lines}. A selection of the features are taken from \cite{Fanelli1992-UVLibrary}. Others were defined by inspection of test model spectra combined with reference to NED\footnote{The NASA/IPAC Extragalactic Database (NED).
is operated by the Jet Propulsion Laboratory, California Institute of Technology,
under contract with the National Aeronautics and Space Administration.} and \cite{NIST_ASD}.
The photometric filters are one of the standard five-band filter sets used in the LEGUS project \citep{Calzetti2015-LEGUSi}: F275W, F336W, and F555W filters from Wide Field Camera 3 (WFC3 UVIS), and the F435W, and F814W filters from the Advanced Camera for Surveys (ACS).
In addition to photometry and equivalent widths, we also generate full spectra for the models referred to as `Sets' 1 to 6.

We note that when they form, star clusters are embedded in gaseous clouds, and are therefore unable to be probed by observations as required by this study. It can take up to a few Myr for these clouds to be cleared, even for low mass clusters \citep[As detailed by, for example 
][]{Hollyhead2015-Gas}. However, in this theoretical study, the simulations at the age-point of $0.1\,{\rm Myr}$ are an illustrative extreme low-age case, and in Appendix~\ref{sec:narrowage} we will consider the effects of narrowing our age range to $> 1~\rm Myr$.

\begin{figure*}
\begin{center}
\begin{tabular}{cc}
\includegraphics[scale=0.42,trim={0.45cm 0 0.45cm 0},clip]{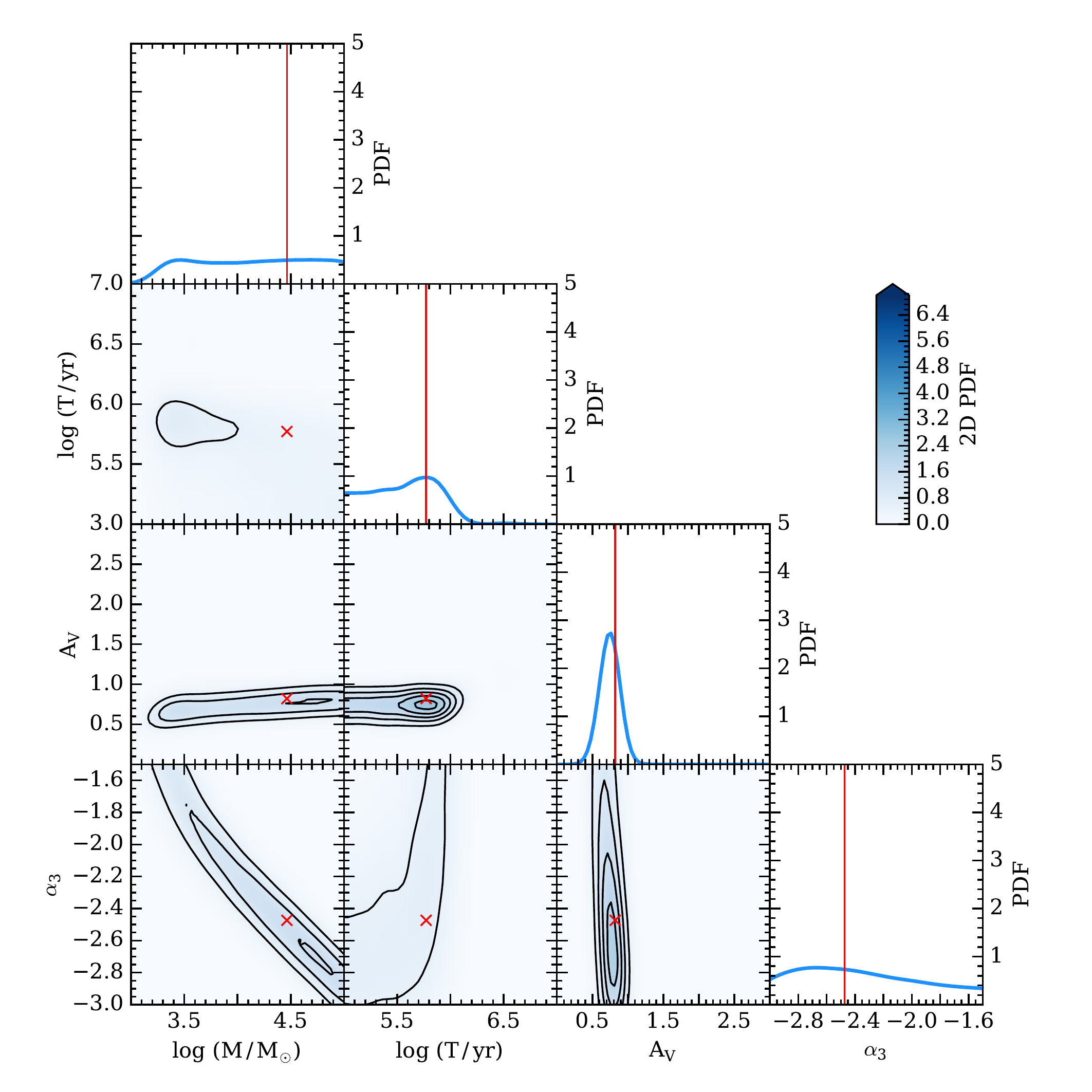} & \includegraphics[scale=0.42,trim={0.45cm 0 0.45cm 0},clip]{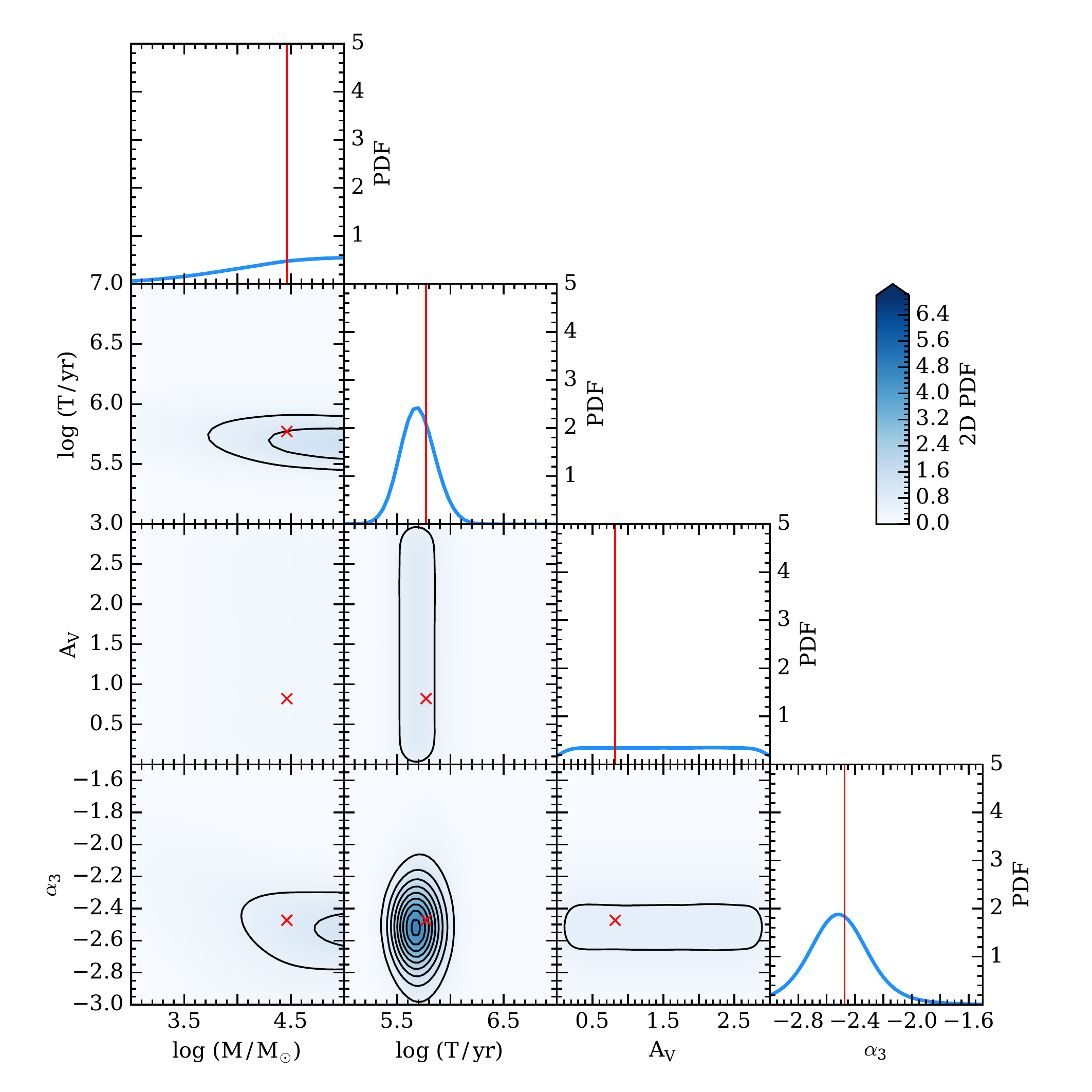} \\
(a) LEGUS Filters & (b) Spectral features: C{\sc iii}, C{\sc iv}a, N{\sc v}a, O{\sc v}a, O{\sc vi}b   \\[6pt]
\includegraphics[scale=0.42,trim={0.45cm 0 0.45cm 0},clip]{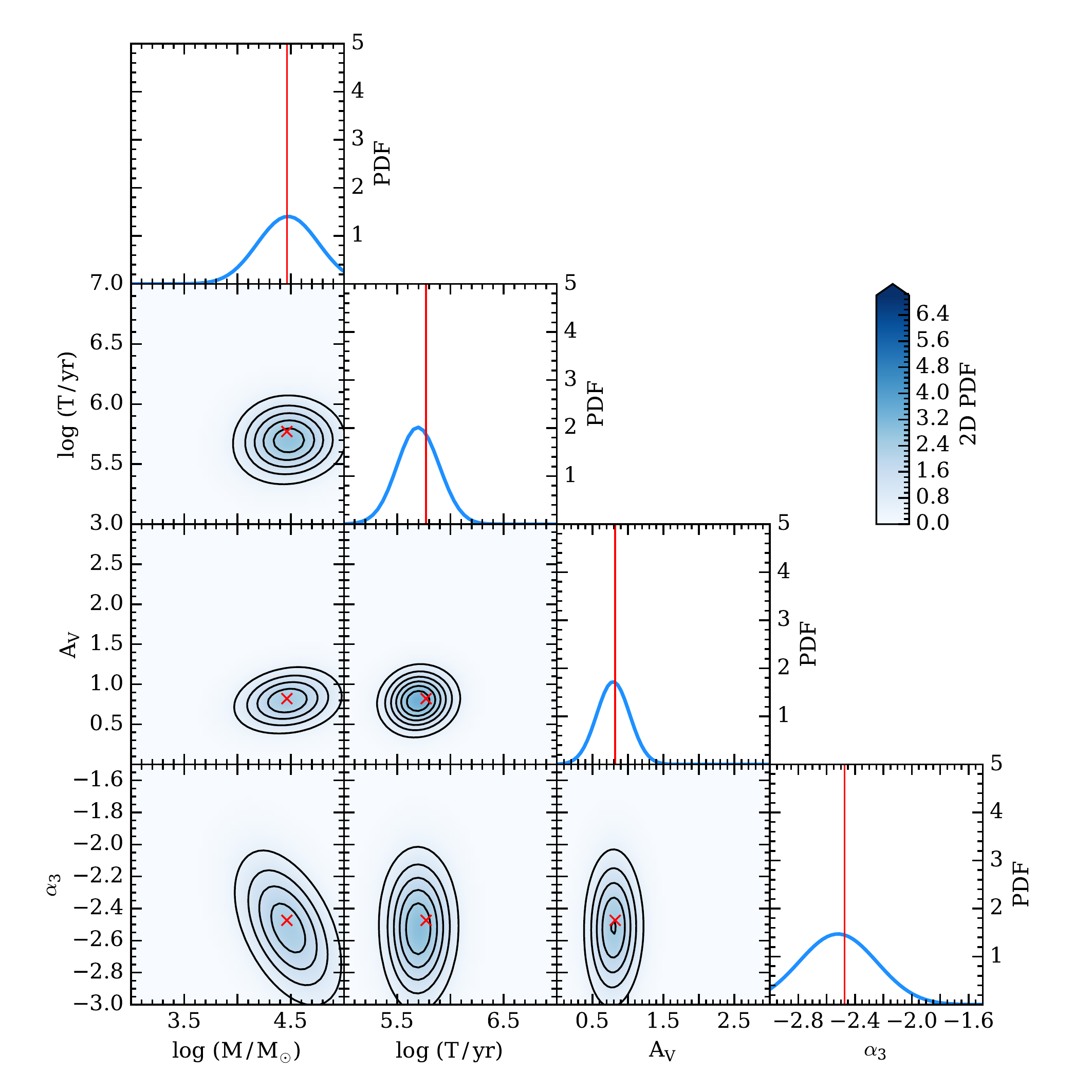} & \includegraphics[scale=0.42,trim={0.45cm 0 0.45cm 0},clip]{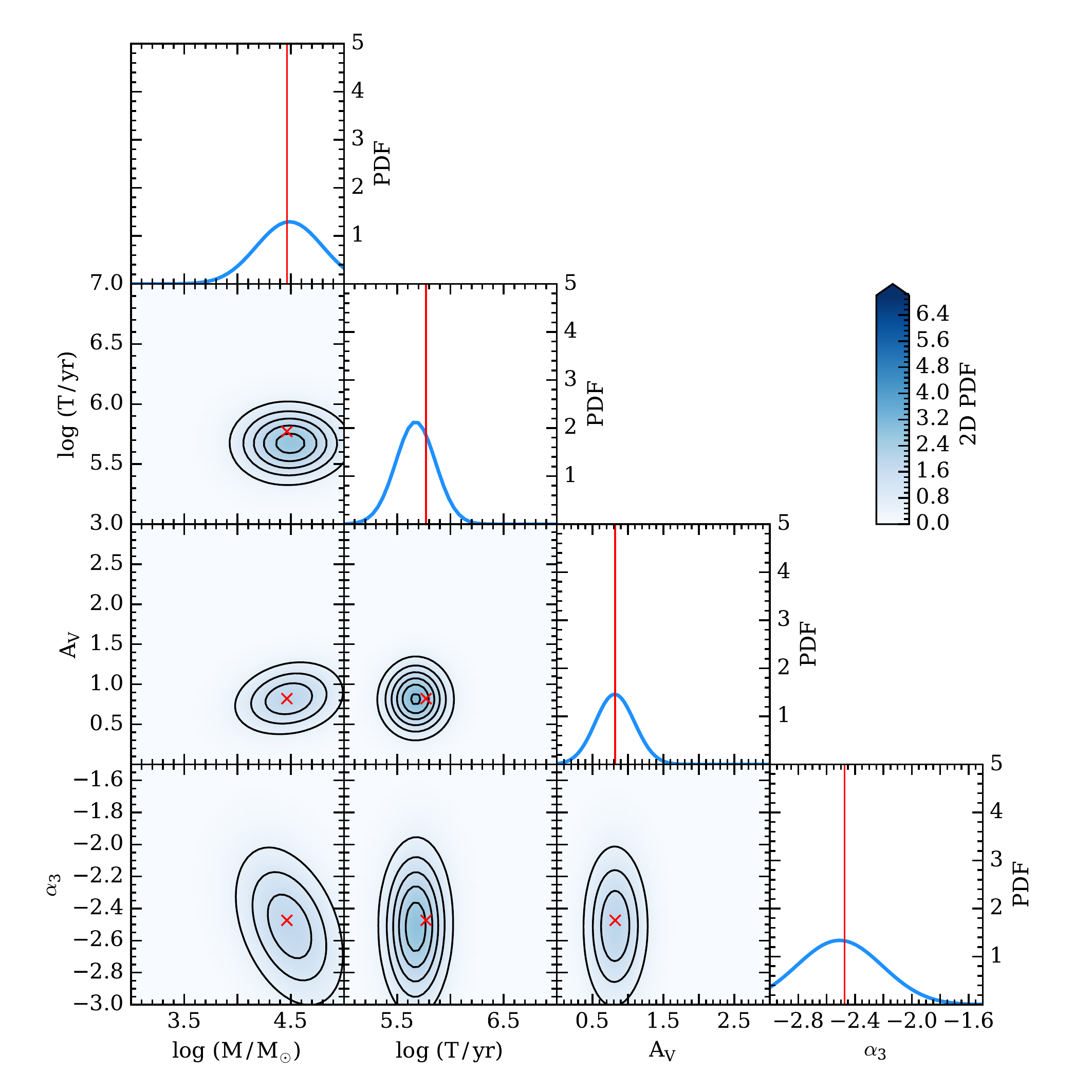}\\
(c) Filters + C{\sc iii}, C{\sc iv}a, N{\sc v}a, O{\sc v}a, O{\sc vi}b   & (d) Filters + All independent spectral features\\[6pt]
\end{tabular}
\end{center}
\centering
\caption{Example corner plot for one of the mock observation clusters, cluster 19. The 1D posterior PDFs for $\log M$, $\log T$, $A_{\rm V}$ and \imfs\ are shown in the top panels of each column. The joint posterior PDFs are represented by contours (beginning at 0, with increments of 0.5 unit element). They are also represented by a 2D map, where the intensity of the colour reflects the probability density as indicated on the colour bar. All PDFs are normalised to have unit integral. In panel (a) are the results when the analysis is performed using only the LEGUS filters, in panel (b) we show results from using the representative selection of lines only, and in panel (c) we show results from using a combination of the two. In panel (d) we present the result from analysis using all the line features that are independent (see Table~\protect\ref{tab:lines}). Note the improvement in the recovery of both the cluster's age and IMF slope when including the spectral lines in the analysis. The red crosses and lines mark the positions of the true values for the parameters that describe this cluster.}
\label{fig:singlecluster}
\end{figure*}

\begin{figure}
\begin{center}
\begin{tabular}{c}
\includegraphics[scale=0.45,trim={0.2cm 0 0.25cm 0},clip]{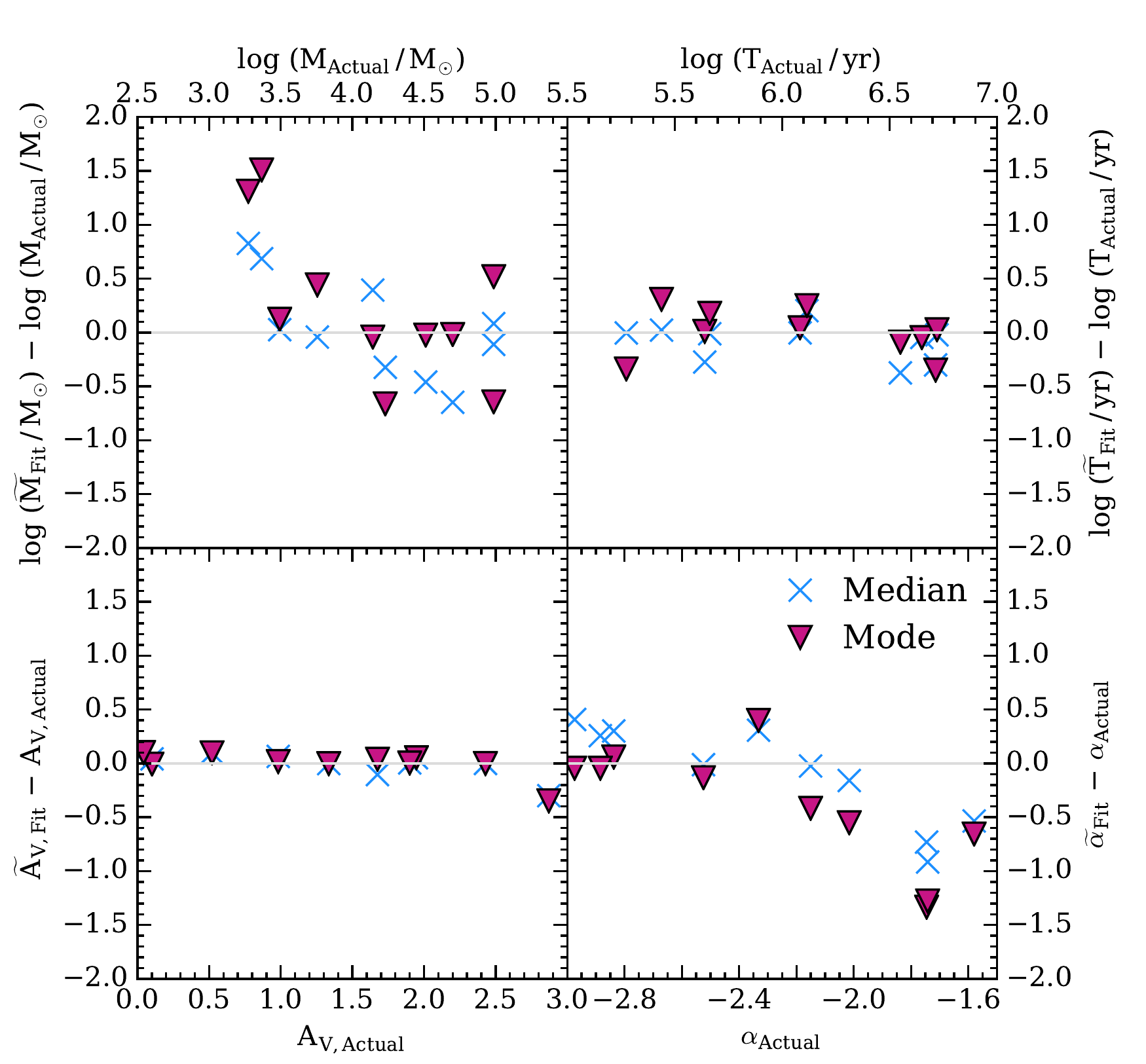}  \\
(a) LEGUS filters \\[6pt]
\includegraphics[scale=0.45,trim={0.2cm 0 0.25cm 0},clip]{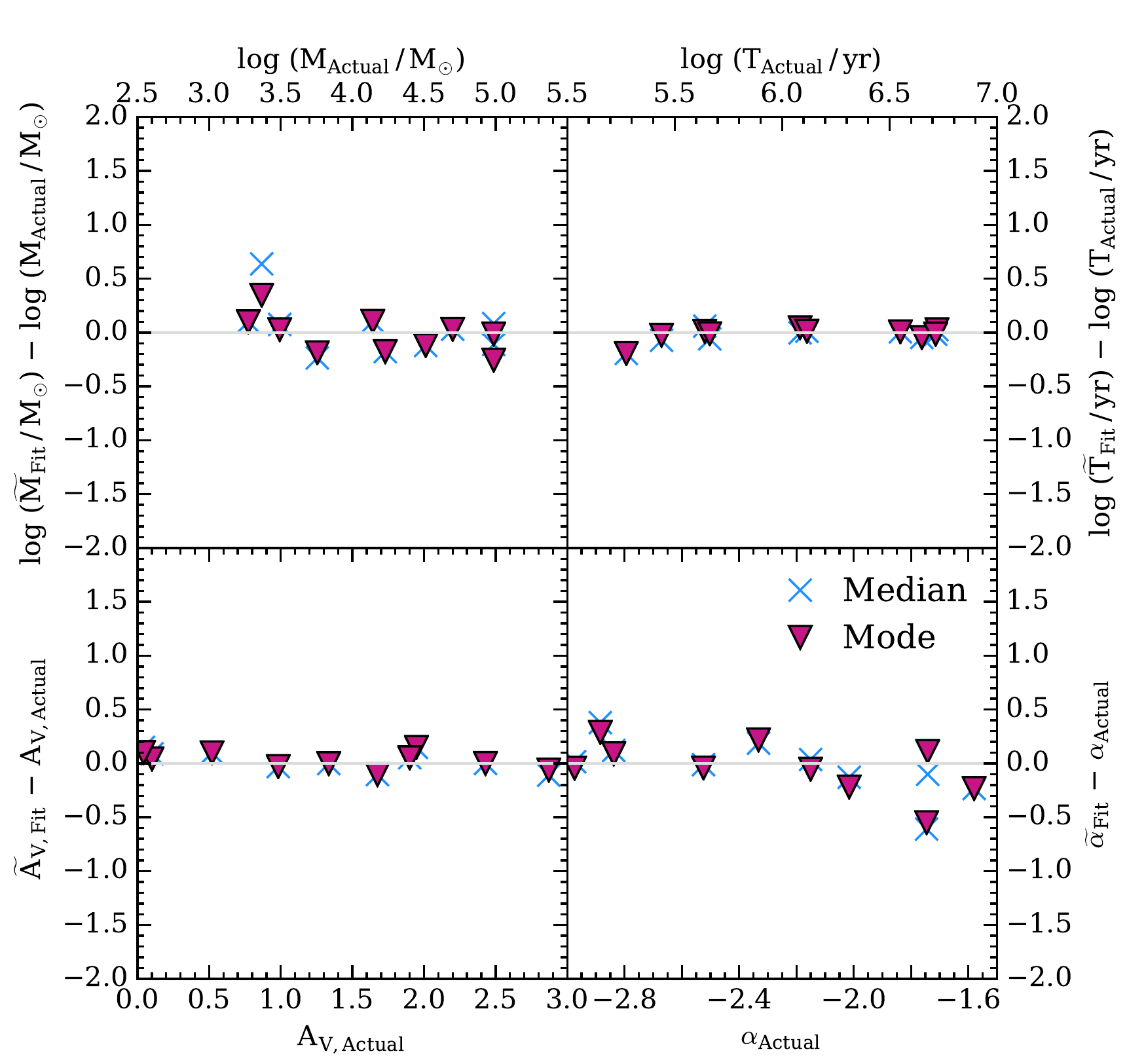}\\
(b) LEGUS filters + lines (C{\sc iii}, C{\sc iv}a, N{\sc v}a, O{\sc v}a, O{\sc vi}b ) \\[6pt]
\end{tabular}
\end{center}
\centering
\caption{Residuals of the 1D posterior PDFs for the parameters $\log M$, $\log t$, $A_{\rm v}$, and \imfs\ for 10 of the clusters generated to provide mock observations. The residuals are calculated from both the medians of the distributions (blue crosses) and the modes of the distributions (purple triangles), when compared to the true values for each of these parameters. We see that the inclusion of the lines into the analysis improves the recovery of all of the parameters, and that in most cases the median and mode move closer together, indicating a narrowing of the distributions as well.}
\label{fig:resid}
\end{figure}

\subsection{Effects of physical parameters on spectral features}\label{sec:modelfeatures}

To investigate the effects of changing the variable physical parameters on star cluster spectra in a qualitative manner, we run several sets of \slug\ models for a variety of IMF slopes, metallicities, cluster masses, and cluster ages. Their parameters are listed in Table~\ref{tab:modelparams} as Sets 1--6.

As a first example, we look into the effects of changing the value of \imfs\ on the equivalent widths of the spectral features, and how these equivalent widths then evolve with time. This is done for both low mass ($10^3\,\msun$) clusters and high mass ($10^6\,\msun$) clusters to get a handle on the effects of stochasticity on the equivalent widths we observe. Results are shown in Figure~\ref{fig:singlefeature} for the N{\sc iv} feature, where the points correspond to the median value from the set of 1000 realisations, and the error bars cover the extent of the 5th to 95th percentile range. We stress that the size of the error bars is not Poisson error, but arises from the stochastic sampling of the IMF (an effect most pronounced in low mass star clusters and clusters with bottom heavy IMFs).

The effects of stochasticity are readily apparent when comparing the low and high mass clusters (panels (a) and (b) to panels (c) and (d) respectively), with the low mass clusters having a wide scatter in equivalent width at all times. This stochasticity is mostly absent for the large $10^6\,\msun$ clusters, with some scatter being apparent for the extremely steep IMF, as expected, due to the small possibility of generating a large star instead of the much more probable large number of small stars.

In the case of the IMF with the $120\,\msun$ cutoff (panels (a) and (c)), we see an upward trend in equivalent width for all three of the \imfs\ values, although it is somewhat obscured by the scatter in the case of the low mass cluster with a steep IMF. We see much the same behaviour for the truncated IMF in panels (b) and (d), with an overall increase in equivalent width compared to the IMF with the higher mass cutoff. This truncation results in a reduction of extremely hot and massive stars, which in turn leads to less of the Nitrogen being highly ionised, and therefore results in more absorption at young ages for shallower IMF slopes.
Already we are able to infer that including lines in our analysis helps to constrain \imfs, although this may be hampered by stochasticity at low masses.

We now move to examine the shape of the spectra themselves, a study which relates to the analysis of stacked spectra. We also look at how metallicity affects the spectrum, and how this compares with the effects caused by variations in the IMF's parameters. 
A sample of the spectrum around the N{\sc v} features is given in Figure~\ref{fig:spectra_dz}, with panel (a) showing how the spectrum varies at a fixed metallicity (solar), fixed age ($10^5\yr$), and fixed mass ($10^6\,\msun$) for changing values of the IMF slope ($\alpha_3=-3,-2.3,-1.5$). Panel (b) shows the how the spectrum varies at the same fixed age and mass, but with a fixed IMF slope of $\alpha_3=-2.3$, and metallicities of $Z=0.004,0.020,0.050$.
In the changing IMF case, we see that the strength of the absorption line feature increases as the IMF becomes more shallow, due to the increased probability of producing hot, massive stars. 
We see that changing metallicity has a similar effect on the N{\sc v} line that changing the IMF does, and a high-metallicity environment appears to be degenerate with a top heavy IMF in the ``N{\sc v}a region'' of the feature. However, the ``N{\sc v}b integration region'' (the emission feature) is not appreciably affected by the changing metallicity, whereas it does differentiate between the three IMF slopes.  

Unfortunately, we are at present unable to vary the metallicity continuously in \slug\ because we lack a set of atmosphere models that is sufficiently densely sampled to allow for effective interpolation.
Should a continuous variation of metallicity be implemented in \slug, it would be possible to introduce it as a fifth physical variable in the Bayesian analysis. This is a degeneracy which should be explored in future work when the necessary features have been implemented in \slug. For the rest of this study, however, we maintain a constant solar metallicity.

Figure~\ref{fig:squares} shows how the equivalent width of two spectral features changes through the age-\imfs\ plane. Each panel contains the same set of $10^6$ cluster realisations drawn randomly from the distributions listed in Table~\ref{tab:modelparams} (the `Squares' set of models). The equivalent width assigned to each grid element is the median of all the realisations that fall within that element. Panel (a) shows the plane for the C{\sc iii} feature, and panel (b) shows the plane for the C{\sc iv}a feature, chosen as they both illustrate well how the equivalent width changes with the two variables, the latter being a strong line and the former being weak.
We see that both C{\sc iii} and C{\sc iv}a allow us to differentiate the IMF slope at young ages, whereas for older clusters we lose this distinguishing power as the massive stars that are responsible for the line features evolve outside the main sequence. Note that, unlike the C{\sc iv} line which increases in strength as the IMF becomes shallow, the C{\sc iii} line becomes weaker as this happens. This effect occurs because a shallower IMF results in a higher proportion of the cluster's light coming from stars that lack the C~\textsc{iii} feature due to them being hot enough to ionise carbon past the C$^{+2}$ state responsible for producing the line. 

We once again note that observations are limited in how young an age they are able to probe, and would be likely limited to ages of a few Myr or greater. However, even at these ages, we still see useful differences in the equivalent widths in Figure~\ref{fig:squares}.

\section{Analysis of mock observations}\label{sec:mockob}

Having discussed the qualitative behaviour of the equivalent width as a function of physical parameters, we now move to study more quantitatively how combining the equivalent widths of spectral features with broad-band photometry may be used to constrain the slope of the IMF, \imfs.

The mock observations are produced from the photometry and equivalent widths of a selection of \slug\ model clusters. These clusters were generated by drawing their physical parameters randomly, as listed in Table~\ref{tab:modelparams} (``Mock Obs.''). It is using this set of mock observations that we attempt to recover the underlying physical properties of the corresponding simulated clusters.
For each set of mock observations we extract the photometry and equivalent widths, assigning each a nominal error of $0.05$. 
We must stress that the mock clusters are not taken from the library, and are generated independently. However, as we use the same tracks and atmospheres to generate the mock clusters, the result is somewhat optimistic compared to observations.

The first step is to generate posterior PDFs (1D and 2D) for the physical parameters of each mock cluster that we have generated. An example (cluster 19) is plotted in Figure~\ref{fig:singlecluster}, with red markings corresponding to the true values of each parameter. The analysis was performed four times for each cluster. Firstly, using one of the LEGUS filter sets \citep[as used for NGC 628e, see][]{Adamo2017-NGC628}, which is comprised of the F275W, F336W, and F555W filters from WFC3 UVIS, and the F435W, and F814W filters from ACS. Next the procedure is repeated using the following spectral features in place of the photometry: C{\sc iii}, C{\sc iv}a, N{\sc v}a, O{\sc v}a, O{\sc vi}b. These features were selected as a representative set of five, with reference to their appearance in the IMF-age plane (c.f. Figure~\ref{fig:squares}), which we use to identify those lines most able to distinguish the cluster age or the IMF slope whilst also giving us a range of line strengths and including an emission feature. Next, the analysis is repeated using both the photometry and the equivalent widths of these spectral features together, first using this representative set of five lines, and then finally using all the spectral features listed in Table~\ref{tab:lines} marked with a $\dagger$ symbol. We cannot include all the line features in the table in our analysis at the same time, as not all the features are independent. Some of our features are sub-features of a single feature we also list, are blends containing many features, or are even multiple occurrences of the same line. For example, we list both the full line feature and the separate components of the lines with P-Cygni profile. For the purposes of this analysis, we choose to use only the absorption component when using the ``full'' independent line list, and indeed the shoulder features may depend greatly on the assumed wind model and therefore may include more uncertainties. We also discard the weaker feature when a line occurs more than once, and we remove the blend features which contain multiple lines.

Panel (a) of Figure~\ref{fig:singlecluster} shows the results of using the broad-band photometry alone. We see quite good recovery of the cluster age and the extinction. However, recovery of the cluster's mass and the value of \imfs\ is poor, due to the underlying degeneracy between the two. Conversely, in panel (b) we see that lines alone recover the age and the IMF slope much more successfully, at the expense of the mass and the extinction. This is expected as $A_{\rm V}$ is unconstrained by construction, and the equivalent width (having been calculated from a normalised spectrum) is weakly dependent on the total cluster mass. In panel (c), we see the combination of the two results in good recovery for all the physical parameters. Indeed, the inclusion of the UV lines further refines our determination of the cluster's age, and helps to alleviate the mass-IMF degeneracy that effects analysis performed with broad-band photometry alone \citep{Ashworth2017-SLUGandLEGUS}, although it does not fully break it. Panel (d) is the same as panel (c), but the analysis was performed with all the line features. The results are mostly comparable, with slight shifts towards more accurate mass, age, and IMF, and a spreading out in the extinction direction. We conclude therefore, that the representative set of five lines is sufficient to greatly improve the recovery of the cluster's physical parameters.

In Figure~\ref{fig:resid} we plot the residuals (the difference between either the median or mode of the posterior PDFs and the true value of the corresponding parameter) for ten of the mock observation clusters.
We see much the same effect as we see in Figure~\ref{fig:singlecluster}  in most cases, and there is some improvement overall in the recovery of the IMF. Note also how the medians and modes move much closer together, as the posterior PDFs become narrower. The overall trends seen for Cluster 19 in Figure~\ref{fig:singlecluster} apply more generally across our mock observations.

\begin{figure*}
\begin{center}
\begin{tabular}{cc}
\includegraphics[scale=0.42,trim={0.45cm 0 0.45cm 0},clip]{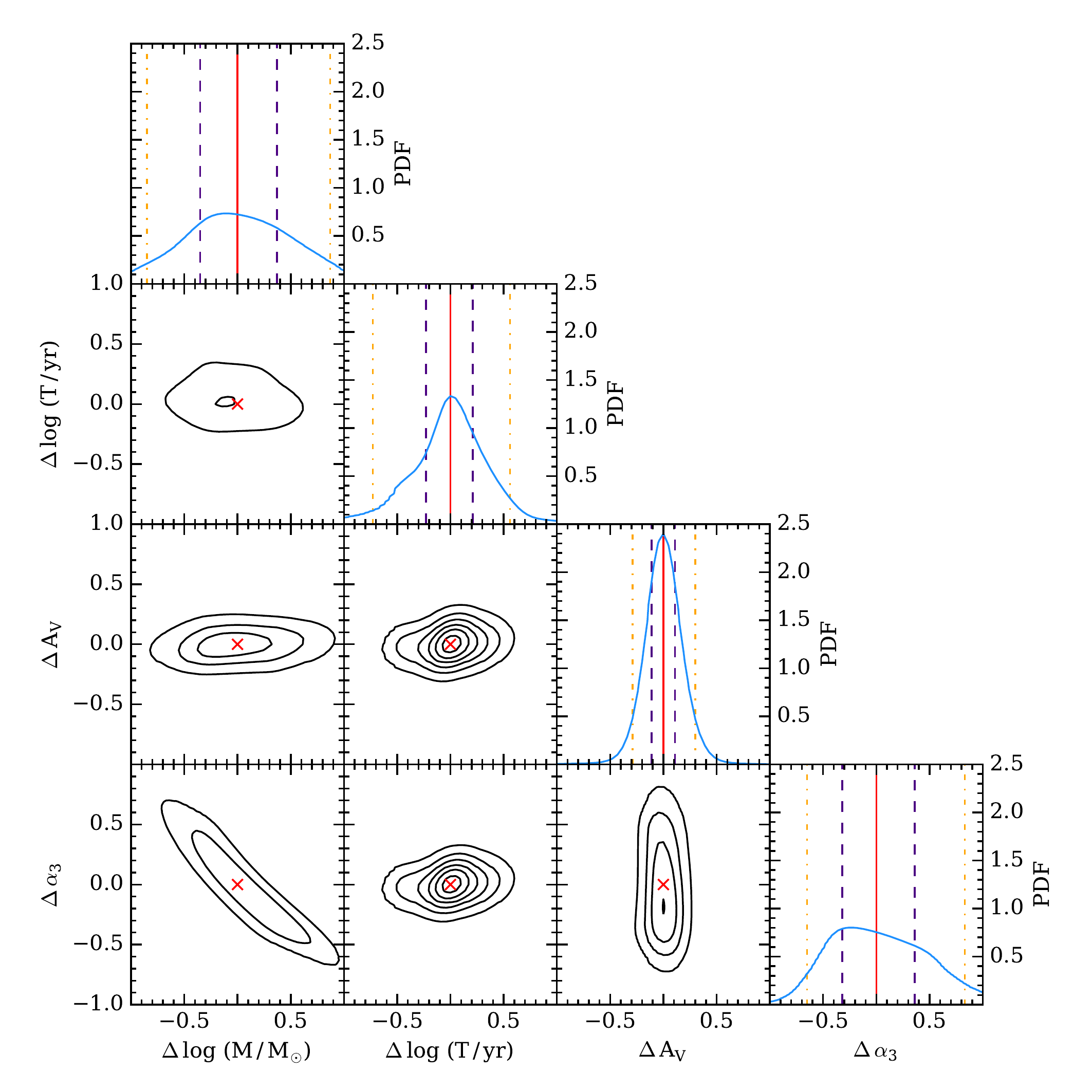} & \includegraphics[scale=0.42,trim={0.45cm 0 0.45cm 0},clip]{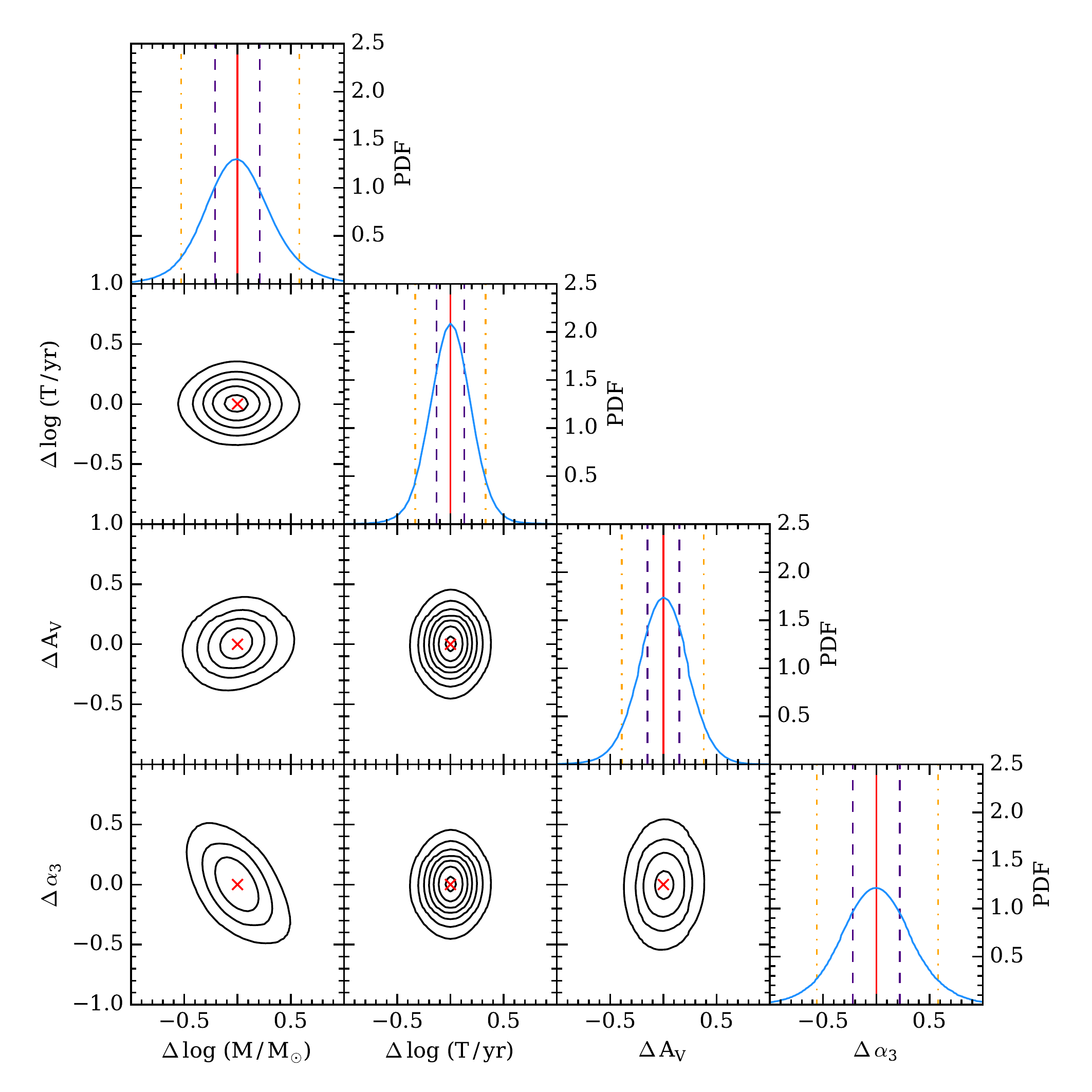} \\
(a) LEGUS Filters, Median centred & (b) Lines and LEGUS filters, Median centred   \\[6pt]
\includegraphics[scale=0.42,trim={0.45cm 0 0.45cm 0},clip]{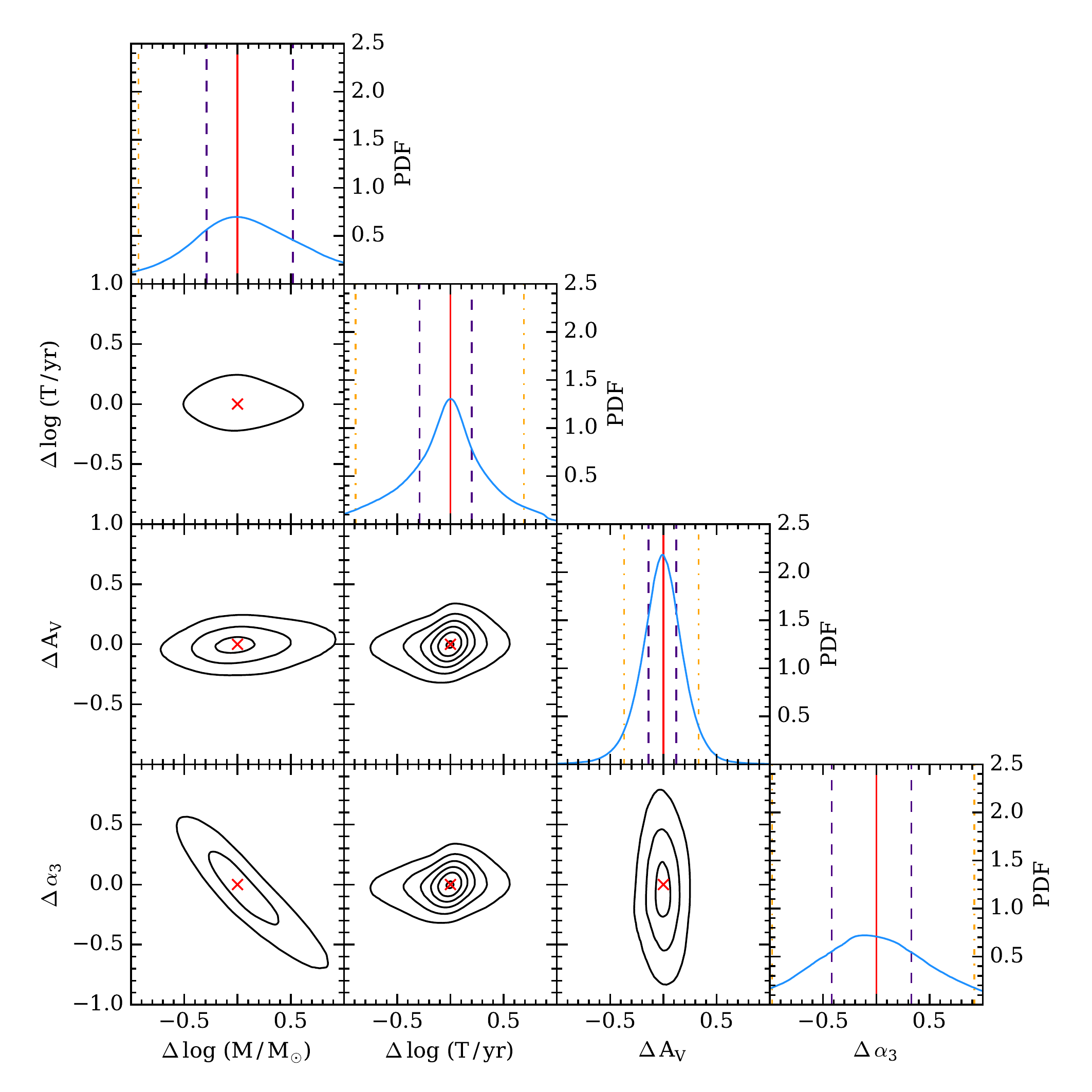} & \includegraphics[scale=0.42,trim={0.45cm 0 0.45cm 0},clip]{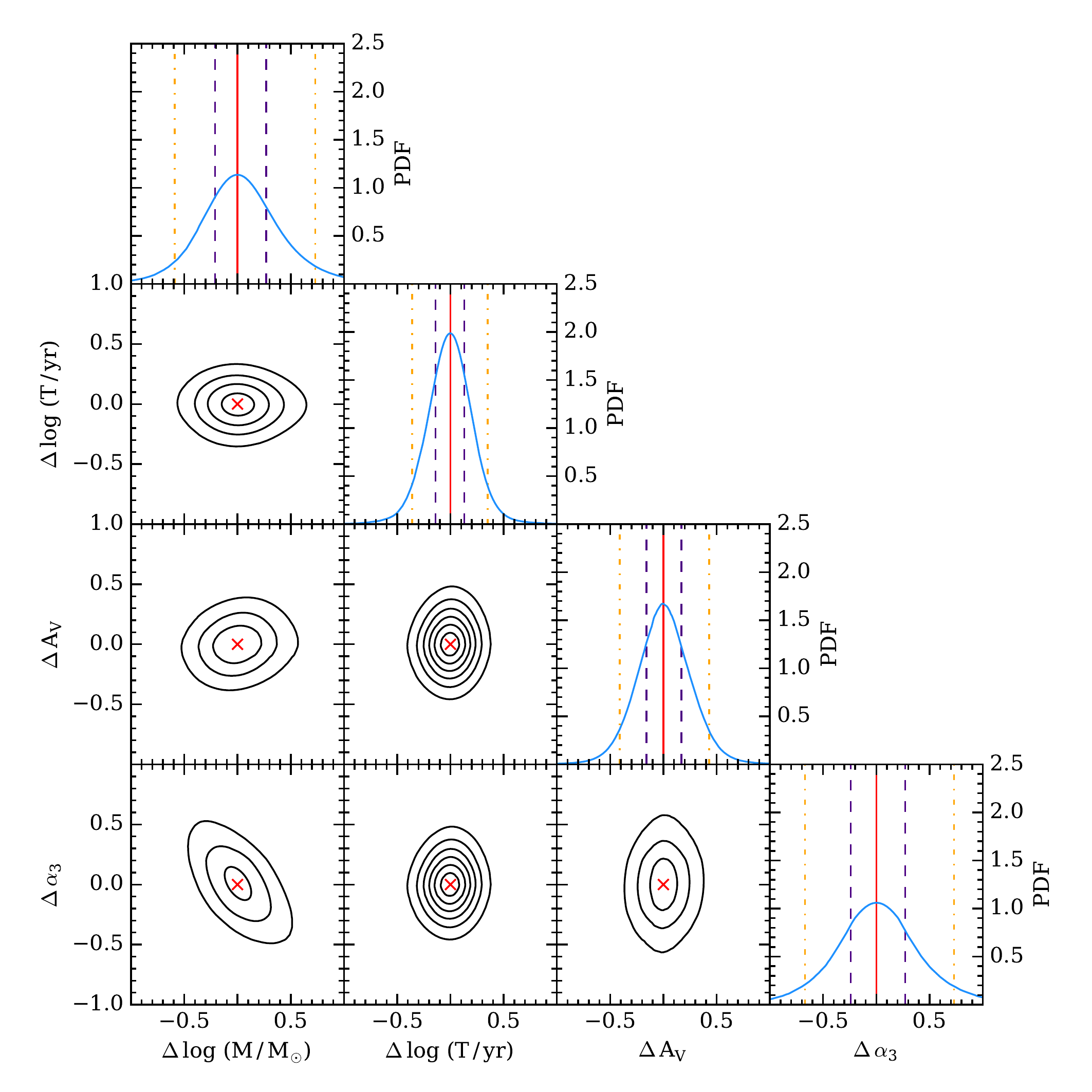}\\
(c) LEGUS Filters, True centred   & (d) Lines and LEGUS filters, True centred \\ [6pt]
\end{tabular}
\end{center}
\centering
\caption{Stacks of the one and two-dimensional PDFs for 500 clusters generated as mock observations, as described in Table~\ref{tab:modelparams}. The stacked PDFs are normalised by the number of clusters being stacked, with contours spaced in steps of 0.5, beginning at 0.5. The red crosses and lines mark the zero points. Panels (a) and (c) are the results of analysis using the LEGUS filters only, and are median-centred and centred on the true values respectively. Likewise, panels (b) and (d) show the results of analysis using both the lines (C{\sc iii}, C{\sc iv}a, N{\sc v}a, O{\sc v}a, O{\sc vi}b) and the LEGUS filters, once again median-centred and true-value-centred respectively. Note the significant improvement in the recovery of the cluster age when including the lines in the analysis. We also see improvement in the recovery of the cluster mass and IMF, although with a residual $\log M$-\imfs\ degeneracy in the joint posterior PDF for those parameters. The fifth and ninety-fifth percentiles are marked with orange dot-dashed lines, and the first and third quartiles are marked with indigo dashed lines. The change in the interpercentile ranges (for the true-centred cases, panels (c) and (d)) are given in Table~\protect\ref{tab:ipr}. }
\label{fig:2dstacks}
\end{figure*}

\input{table_stacks.tex}

\begin{figure*}
\begin{center}
\begin{tabular}{cc}
\includegraphics[scale=0.505,trim={0.43cm 0 0.4cm 0},clip]{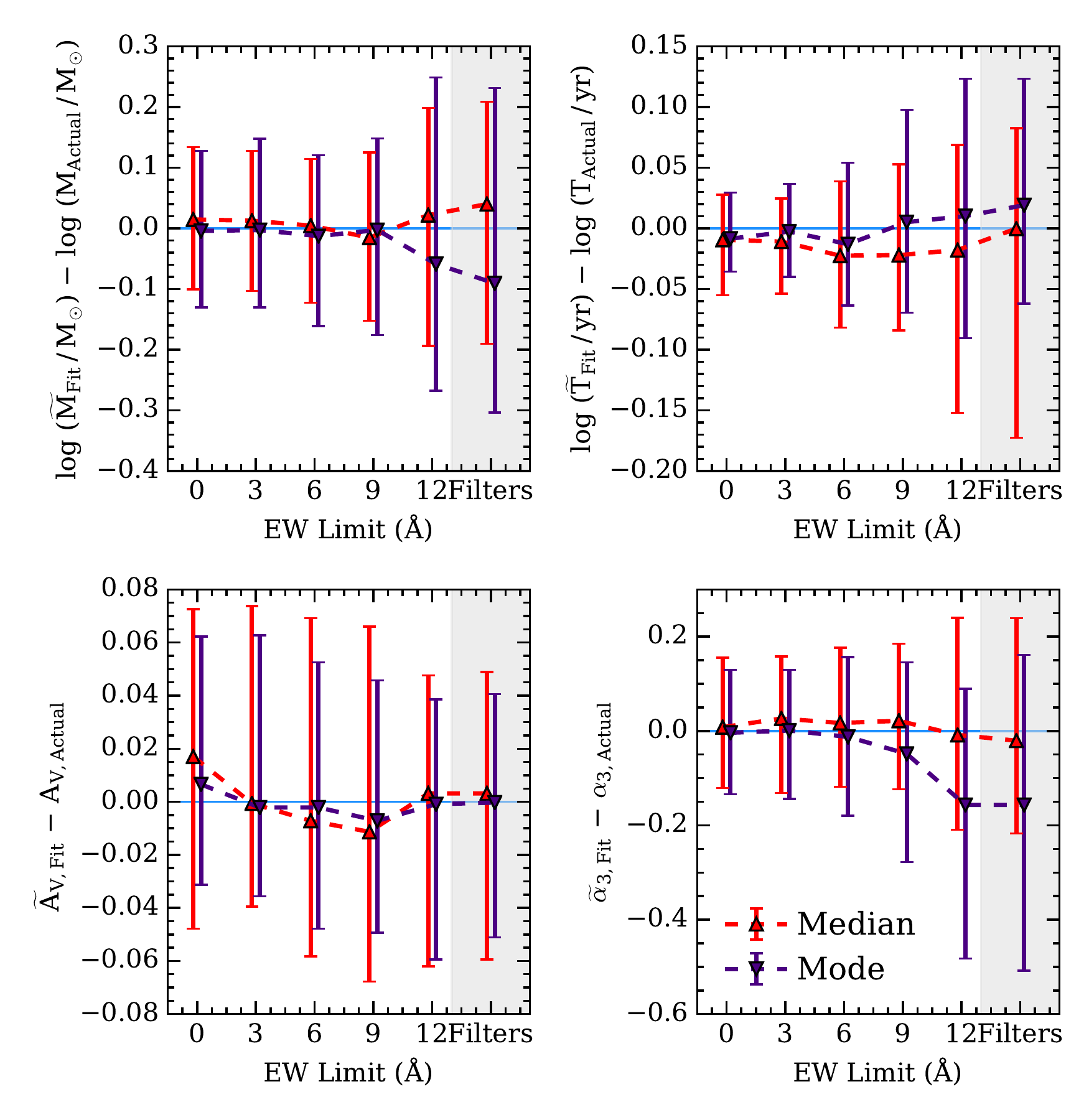} &
\includegraphics[scale=0.505,trim={0.43cm 0 0.4cm 0},clip]{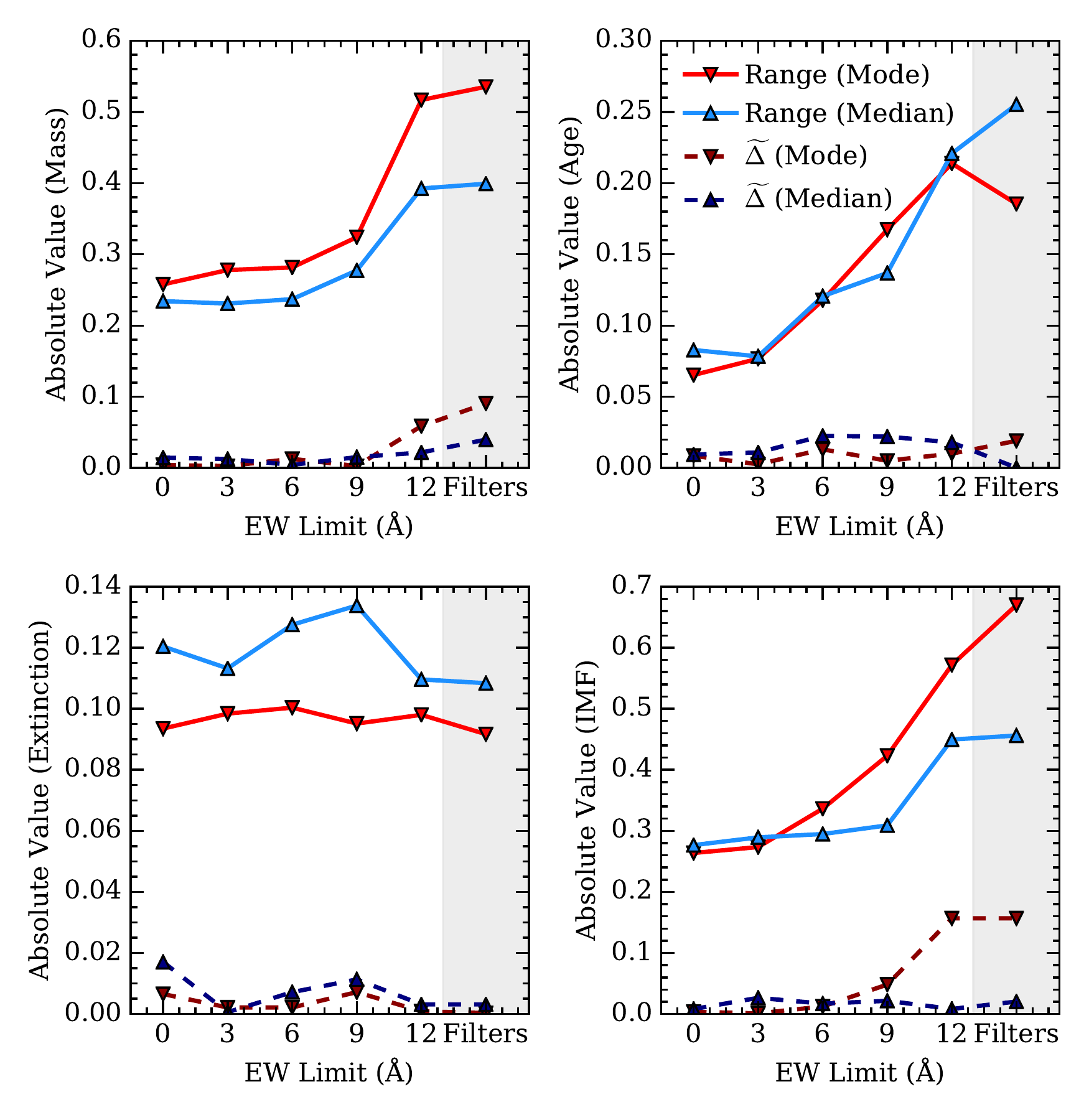} \\
(a) & (b)\\[6pt]

\end{tabular}
\end{center}
\caption{The plots in panel (a) show the residuals of the 1D posterior PDFs for the parameters $\log M$, $\log t$, $A_{\rm v}$, and \imfs\ for the first 100 mock observation clusters. The residuals are calculated from both the medians of the distributions (red upward-pointing markers) and the modes of the distributions (purple downward-pointing markers), when compared to the true values for each of these parameters. 
For each cluster we have set a value of equivalent width below which no spectral features are included in the analysis.
The markers correspond to the median value of the residual at each equivalent width limit, and the error bars cover the interquartile range. The markers are offset by $\pm 0.2$ around their limit value for visual clarity.
Panel (b) illustrates the behaviour of both the interquartile range, and the absolute value of the median residual for both the case where it is calculated from the modes of the PDFs (reds, downward triangles) and the medians (blues, upward triangles). The markers in panel (b) are not offset, as there are no errorbars to overlap.}
\label{fig:ewlimsresid}
\end{figure*}

Finally, we quantify the overall behaviour of the PDFs across our entire input dataset by looking at stacks of the all the posterior PDFs for 500 mock observation clusters, which we show in Figure~\ref{fig:2dstacks}. In panels (a) and (b) the posterior PDFs are centred on their medians, whereas those in panels (c) and (d) are centred on the true values of each parameters for each of the clusters in the stack. Panels (a) and (c) are produced using only the five LEGUS filters in the analysis, whereas panels (b) and (d) are produced using a combination of the both the photometry and the equivalent widths of the lines in our representative set of five. The 5th and 95th percentiles are marked with orange dot-dashed lines, and the first and third quartiles are marked with indigo dashed lines. The zero marker, on which the PDFs are centred as they are stacked, is the solid red line (or cross in the 2D PDFs). 

As seen in the previous examples, there is significant improvement overall in our recovery of the physical parameters once the spectral features are included. Indeed, we see for the true-value centred stacks that we move from quite wide distributions around the value of interest to quite narrow peaks. This can be seen more clearly by looking at the changes in the interpercentile (fifth to ninety-fifth) and interquartile (first to third) ranges, as listed in Table~\ref{tab:ipr} for both the true-centred and median-centred cases. While the median-centred case (marked MC) is a useful companion to panels (a) and (b) of Figure~\ref{fig:2dstacks}, the true-centred rows (marked TC) illustrate the improvement (or worsening) of our recovery of the parameters.  In effect, we see a reduction in the overal scatter (as represented by the 5-95 range) across all parameters, save for the extinction, with improvements having a mean of $\sim25\%$. For the interquartile range, we see a mean reduction in width of $\sim23\%$, with a narrowing of $\sim32\%$ for \imfs.
The broadening of the posterior PDFs for extinction is caused by the lack of extinction information in the rectified spectra we use for calculating the equivalent widths of our spectral features. This results in the almost flat posterior PDFs that we see in Figure~\ref{fig:singlecluster}. It is the addition of the lines, which are insensitive to the extinction, to the extinction sensitive photometry which causes the spreading out of our distributions here. 

While we are unable to entirely break the mass-IMF degeneracy described by \cite{Ashworth2017-SLUGandLEGUS} (curved contours in the joint $\log M$-\imfs\ PDF are still visible even with the equivalent widths included in the analysis), we are able to make significant improvements on constraining the value of not only \imfs, but the mass and age of the clusters as well. We also note that some residual scatter is intrinsic due to the stochasticity in our models, and cannot be reduced by adding more clusters. Indeed, during testing we found no significant difference in scatter between stacks of 10, 100, or 500 clusters for a fixed Kroupa IMF case.
Our analysis has therefore shown, when introducing additional information in the form of UV spectral lines, that it is possible to tighten the constraints on \imfs, and therefore possible to rule out extreme variations in \imfs\ as well.

\subsection{Observational constraints}\label{ssec:obcon}

To aid the planning of new observations, in this section we aim to predict what spectral features are required to gain sufficient improvement in recovery of the IMF when compared to the case of using pure broad-band photometry. What is of interest is specifically what information is needed to get the improved recovery of \imfs\ that we desire. To this end, we take the empirical approach of adding in features by their overall strength (equivalent width) until there is a point of diminishing returns, where the addition of further spectral features does not improve the posterior PDFs by an appreciable amount. 

The most basic method by which we can do this is to limit which spectral features we include in our Bayesian analysis by imposing limits on the equivalent width to which observations may be sensitive. This is a simple approach which allows us, to first approximation, to gauge the sensitivity required by real observations, and thus the wavelength coverage, resolution, and exposure time needed to constrain \imfs\ in a given sample of clusters.

To this end we select 100 example clusters from the `Mock Obs.' model set, and repeat the Bayesian analysis presented in Section~\ref{sec:mockob}. However, instead of using a fixed selection of filters for the entire dataset, we limit the inclusion of spectral features to those which have an equivalent width that passes a particular threshold for each cluster individually. We choose limits of $12,9,6,3\,\ang$, along with the case where we have all the line features included, and the case where we use broad-band photometry alone.
In the case of the $9\,\ang$ limit, this gives us between zero and three line features across the input clusters. 

In panel (a) of Figure~\ref{fig:ewlimsresid}, we see the residuals of the 1D posterior PDFs for the four physical parameters for this set of input clusters. The markers represent the median residual value for each equivalent width limit, with the error bars covering the interquartile range. The markers in red represent the residuals calculated from the difference between the median of the posterior PDFs and the true values, whereas the purple points represent the residuals calculated using the difference between the mode of the posterior PDFs and the true values. 

We note that even at a sensitivity limit of $9\,\ang$, we see definite improvement in age, mass, and IMF determination. As sensitivity increases, we see a slow improvement in recovery for these parameters, although the median residuals are close to zero across the range. However, the residuals for the extinction increase. This is caused by the line features having no extinction applied to them, resulting in an almost flat PDF for the $A_{\rm V}$ parameter as seen in Figure~\ref{fig:singlecluster}.
As more and more spectral features are included, the posterior PDF for extinction is spread out from the narrow peak we obtain when using filters alone, as more and more noisy PDFs are combined with it.
Over this sample of 100 clusters, we find that even with a limit of $9\,\ang$ we are able to significantly improve the accuracy at which we are able to infer the value of \imfs, and indeed that further improvements in sensitivity provide limited improvement over this. In both age and cluster mass, however, further increases in sensitivity continue to visibly improve results.

The improvement in our ability to constrain the IMF is more visible in panel (b) of Figure~\ref{fig:ewlimsresid}. In this figure we see the evolution of the absolute value of the median of the residuals, as well as the evolution of the interquartile range, as we increase the sensitivity. Looking at the interquartile ranges, we see that our overall ability to constrain the IMF improves significantly with the introduction of equivalent width data, with improvement in \imfs\ being $\sim 32\%$ (median residual) or $\sim 37\%$ (mode residual)  at the $9\,\ang$ limit. However, further improvements gained by increasing our sensitivity are small, finally coming to $\sim 39\%$ (median residual) or $\sim 61\%$ (mode residual) when we include all the spectral features in the analysis. We see similar behaviour in mass, whereas the improvement in age is more consistent across the range of equivalent width limits. Our constraints on extinction loosen, for the reason explained earlier.


\section{Summary and conclusions}\label{sec:conclusions}
In this paper we have developed an extension to the SPS code \slug, whereby the resolution of the output spectra has been greatly increased in the UV region (from $\simeq10\,\ang$ to $\simeq0.4\ang$). Also, a new equivalent width calculation mechanism has been implemented to take advantage of these new higher resolution spectra. These new capabilities were then leveraged to perform a theoretical study of the high mass end of the IMF and its relationship to observable quantities for unresolved stellar populations.

Having performed a qualitative exploration of how the equivalent widths of spectral features are affected by changes in the various physical parameters ($\log M$, $\log t$, $A_{\rm V}$, \imfs) that describe our model star clusters, we applied the Bayesian analysis capabilities of {\sc bayesphot} to mock observations generated with \slug. These observations (both broad-band photometry and the equivalent widths of UV line features) were analysed using a large library of \slug\ models which covered a wide parameter space. We found similar difficulty in tightly recovering the value of \imfs\ as in \cite{Ashworth2017-SLUGandLEGUS} where degeneracy between the cluster mass and the IMF slope hampered this recovery.
Repeating the analysis while including the equivalent widths of five representative spectral features resulted in a significant improvement in the recovery of the physical parameters for the clusters when averaged over the whole set of 500 mock observations, with the interquartile range of the IMF posterior PDF shrinking by 32\%. However, we still see evidence of the mass-IMF degeneracy noted in previous work.

Finally, we provided predictions for the required sensitivity in equivalent width needed to make improvements in recovery of \imfs\ over using broad-band photometry alone. We find that a modest limit of $9\,\ang$ provides sufficient improvement (with the interquartile range of the median-residuals shrinking by $32\%$ for the IMF), with further improvement as more line features are included in the analysis. 

The study presented in this work is purely theoretical, and as such provides an idealised situation for recovering the IMF from spectra. Although, due to \slug's stochastic nature, it is not simply a trivial problem of finding the exact match in the library, it would be instructive to attempt to recover the IMF for clusters with an independently calculated \imfs\ value. The real-world performance improvements of the method presented in that study will depend on how well the UV features of the spectrum are modelled. High resolution UV spectral observations of a full resolved stellar population, the central cluster of 30 Doradus \citep{Crowther2016-30Dor} for example, would provide suitable input equivalent widths for Bayesian analysis with \slug, whilst also allowing an independent calculation of the IMF through star counts. 

A limitation of the models generated during this work is the absence of binary systems in the star clusters we build. Binary star systems are thought to make up a large proportion of the star systems present in galaxies, with perhaps all O and B stars residing in binary systems, as summarised by \cite{Eldridge2017-BPASS}. Indeed, the presence of binaries affects the evolution of a given stellar system due to events such as mergers and mass transfer between the two stars, and their presence results in a given population of stars appearing more blue than would be predicted by models that do not contain binaries \citep{Eldridge2012-StochBin}.
The effect of mergers and mass transfer would be to push the IMF to a shallower slope, as massive stars would become even more massive with time as they accrete. In addition, \cite{Eldridge2009-Binaries} note that the inclusion of binaries can spread the Wolf-Rayet phase out across a longer age range, which would affect our analysis if all stellar types were included in the equivalent width calculation. Although capturing these effects in \slug\ requires tracking multiple new parameters (which could be varied within our Bayesian framework at the expenses of computational efficiency), this extension may provide a more realistic library that can be used to constrain the IMF. 
Another limitation is the absence of stellar rotation in our models. This is due to the use of the Padova tacks with thermally pulsating AGB stars which do not include stellar rotation. However, as \slug\ is distributed with the Geneva stellar tracks \citep{genevarot}, which do include the effects of rotation, it would be possible to study what effects (if any) including rotation has on the IMF slopes that we recover. The inclusion of rotation should result in an increase in the luminosity of O stars \citep{Vasquez2007-Rotation,Levesque2012-Rotation}, although the effect of this would be dependent on the speed of this rotation. Once more this would require the addition of further variables into our analysis framework.

In conclusion, we find that the inclusion of UV spectral features does improve recovery of the physical parameters of unresolved stellar populations in our theoretical study, over that possible with broad-band photometry alone. However, we do not find that it makes it possible to fully break the mass-IMF degeneracy that exists in these systems. Having applied cuts to the included line features, we find observations sensitive to a modest equivalent width are capable of improving the recovery of the IMF high-mass slope \imfs\ to a significant degree, with a scatter narrow enough to either rule out or indeed distinguish extreme variations in the IMF.

\section*{Acknowledgements}

G.A. acknowledges support from the Science and Technology Facilities Council (ST/P000541/1 and ST/M503472/1).
M.F. acknowledges support by the Science and Technology Facilities Council (ST/P000541/1). This project has received funding from the European Research Council
(ERC) under the European Union's Horizon 2020 research and innovation
programme (grant agreement No 757535)
A.A. acknowledges the support of the Swedish Research Council (Vetenskapsr\aa det) and the Swedish National Space Board (SNSB).
We also thank the LEGUS Collaboration, along with Claus Leitherer and Wolf-Rainer Hamann for helpful discussions. We wish to thank the referee, Prof. Nate Bastian, for insightful comments on this manuscript. 
This work used the DiRAC Data Centric system at Durham University, 
operated by the Institute for Computational Cosmology on behalf of the 
STFC DiRAC HPC Facility (www.dirac.ac.uk). This equipment was funded by 
BIS National E-infrastructure capital grant ST/K00042X/1, STFC capital 
grants ST/H008519/1 and ST/K00087X/1, STFC DiRAC Operations grant 
ST/K003267/1 and Durham University. DiRAC is part of the National 
E-Infrastructure.
This research has made use of the NASA/IPAC Extragalactic Database (NED),
which is operated by the Jet Propulsion Laboratory, California Institute of Technology,
under contract with the National Aeronautics and Space Administration.
For access to the data and codes used in this work, please contact the authors or visit \url{www.slugsps.com}.

\bibliographystyle{mnras}
\bibliography{paper3bib}

\appendix
\section{Narrowing the age coverage}\label{sec:narrowage}

\input{table_models_2.tex}
\input{table_stacks2.tex}
As noted in Section~\ref{sec:models}, our study covers a wider range in ages than would be generally visible to observations, as stars younger than a few Myr are likely to be still in an embedded phase.
In addition, it is possible that the assumption that the stars in the stellar population are coeval may break down at ages as young as $10^5\,\yr$. For example, \cite{Kudryavtseva2012-StarburstMC} report age spreads in young star clusters which are at least as large as this. 
As this study is purely theoretical, this wide age range is still instructive in understanding the effect of the IMF on the output spectrum at its most extreme. However, as observations may generally not be able to probe ages as young as $10^5\,\yr$, we now repeat our analysis over a more constrained range of ages of $10^{6-7}\,yr$. The parameters that define the clusters used to generate the new library and mock observations are given in Table~\ref{tab:modelparams2}.

\begin{figure*}
\begin{center}
\footnotesize
\begin{tabular}{cc}
\includegraphics[scale=0.35,trim={0.8cm 0 0.85cm 0},clip]{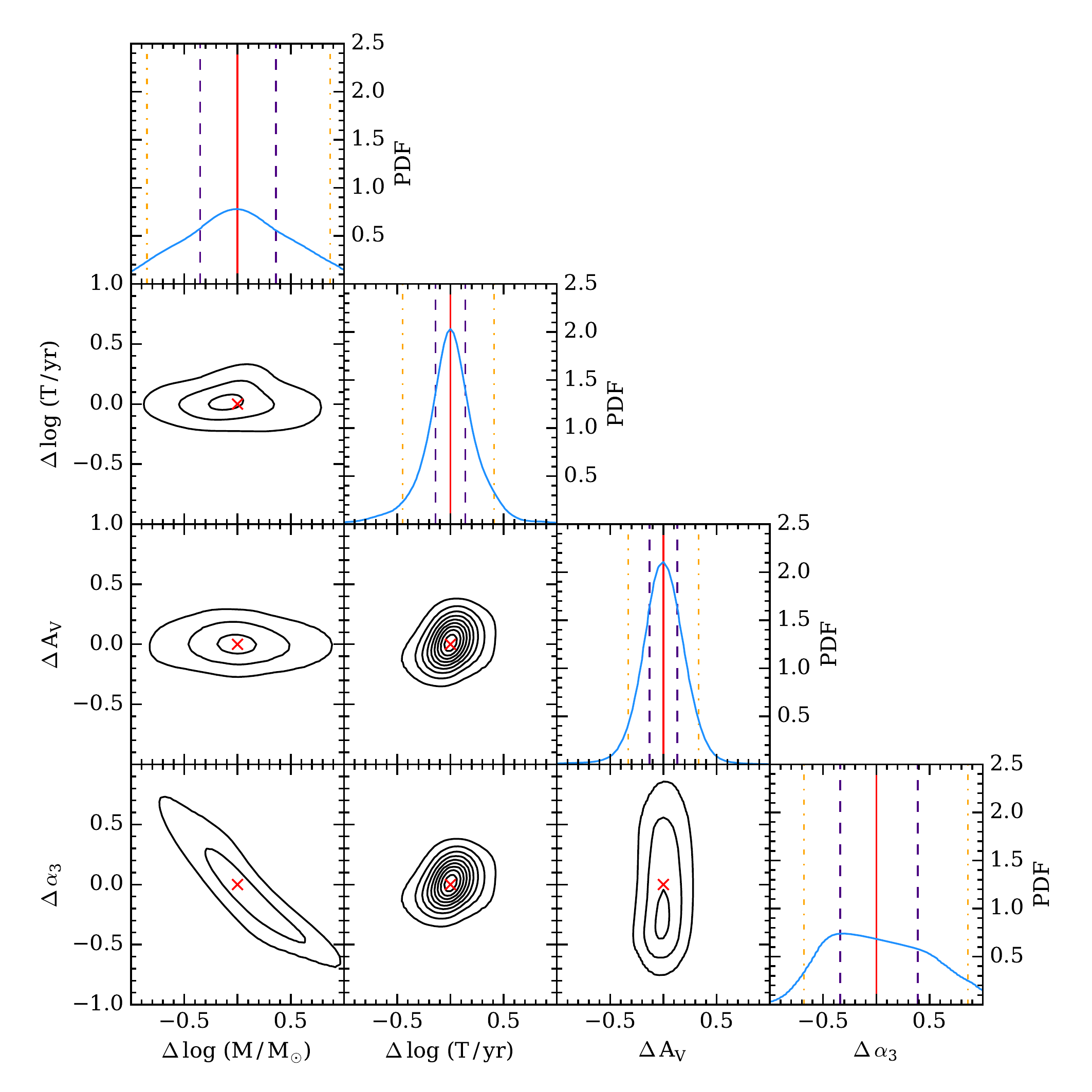} & \includegraphics[scale=0.35,trim={0.8cm 0 0.85cm 0},clip]{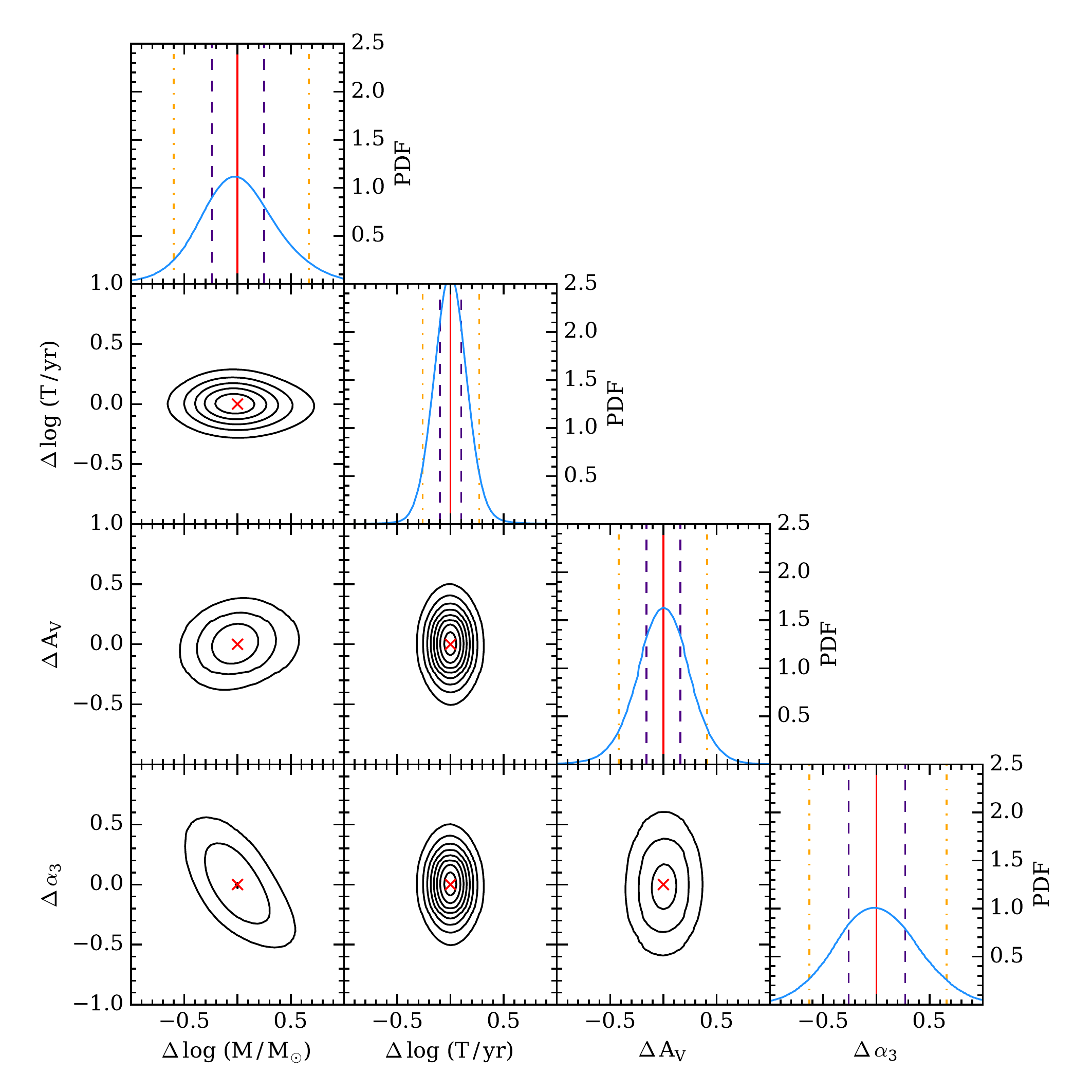} \\
(a) LEGUS Filters, Median centred & (b) Lines + LEGUS filters, Median centred   \\[6pt]
\includegraphics[scale=0.35,trim={0.8cm 0 0.85cm 0},clip]{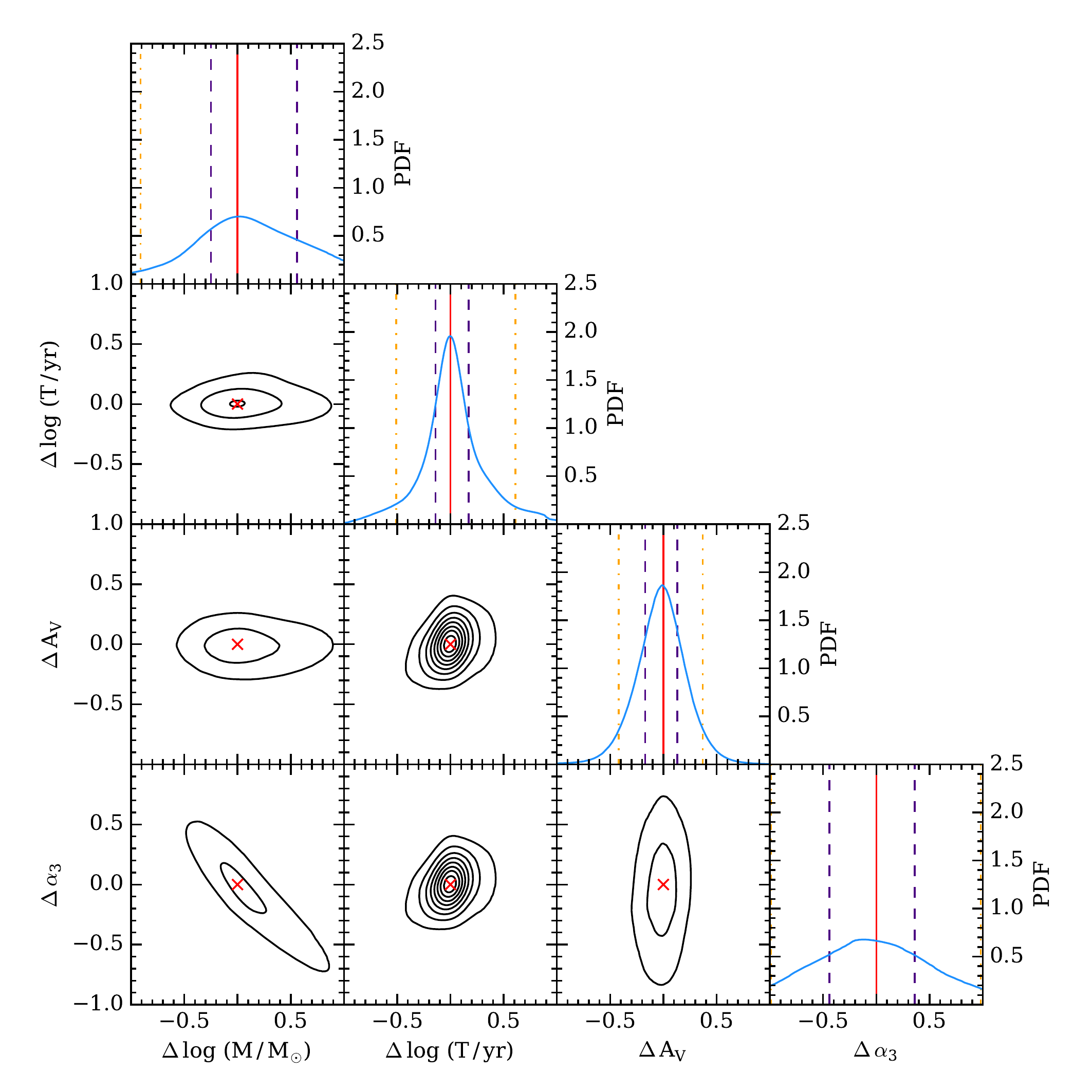} & \includegraphics[scale=0.35,trim={0.8cm 0 0.85cm 0},clip]{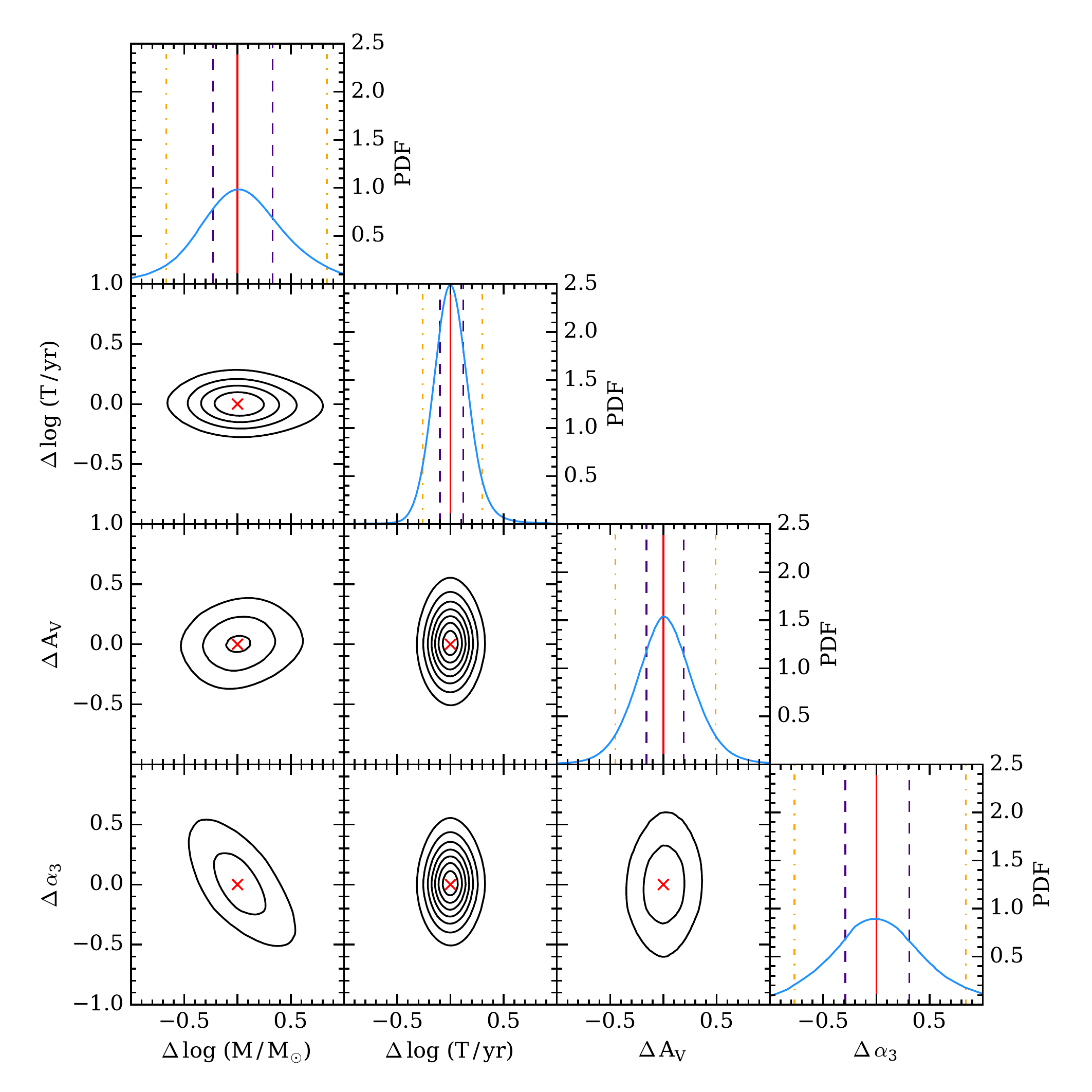}\\
(c) LEGUS Filters, True centred   & (d) Lines + LEGUS filters, True centred \\ [6pt]
\end{tabular}
\end{center}
\centering
\caption{As Figure~\protect\ref{fig:2dstacks} but for the narrower age-range models detailed in Table~\protect\ref{tab:modelparams2}. The change in the interpercentile ranges (for the true-centred cases, panels (c) and (d)) are given in Table~\protect\ref{tab:ipr2}. }
\label{fig:2dstacks2}
\end{figure*}

As before, we generate 500 sets of mock observations with both photometry (across the same LEGUS filterset) and equivalent widths (for the lines listed in Table~\ref{tab:lines}). In Figure~\ref{fig:2dstacks2}, we see very similar behaviour to that seen in Figure~\ref{fig:2dstacks}, with the changes in the interquartile and interpercentile (5-95) ranges given in Table~\ref{tab:ipr2}. The relative improvement in recovery of \imfs\ when restricting ourselves to clusters in the $1-10$\,Myr range is smaller than we saw in Table~\ref{tab:ipr}, with the stacked PDFs in the true-centred case narrowing their interquartile ranges by $25\%$ compared to $32\%$ when including the younger clusters. In addition, apart from the age paramater (where we have effictively applied a more narrow prior), the PDFs are slightly broader than we saw in Figure~\ref{fig:2dstacks}. Even so, we still see that for clusters in the age range likely to be seen by observations, we are able to produce an improvement in IMF recovery of $25\%$ when using the representative set of 5 lines.

\begin{figure*}
\begin{center}
\begin{tabular}{cc}
\includegraphics[scale=0.505,trim={0.43cm 0 0.4cm 0},clip]{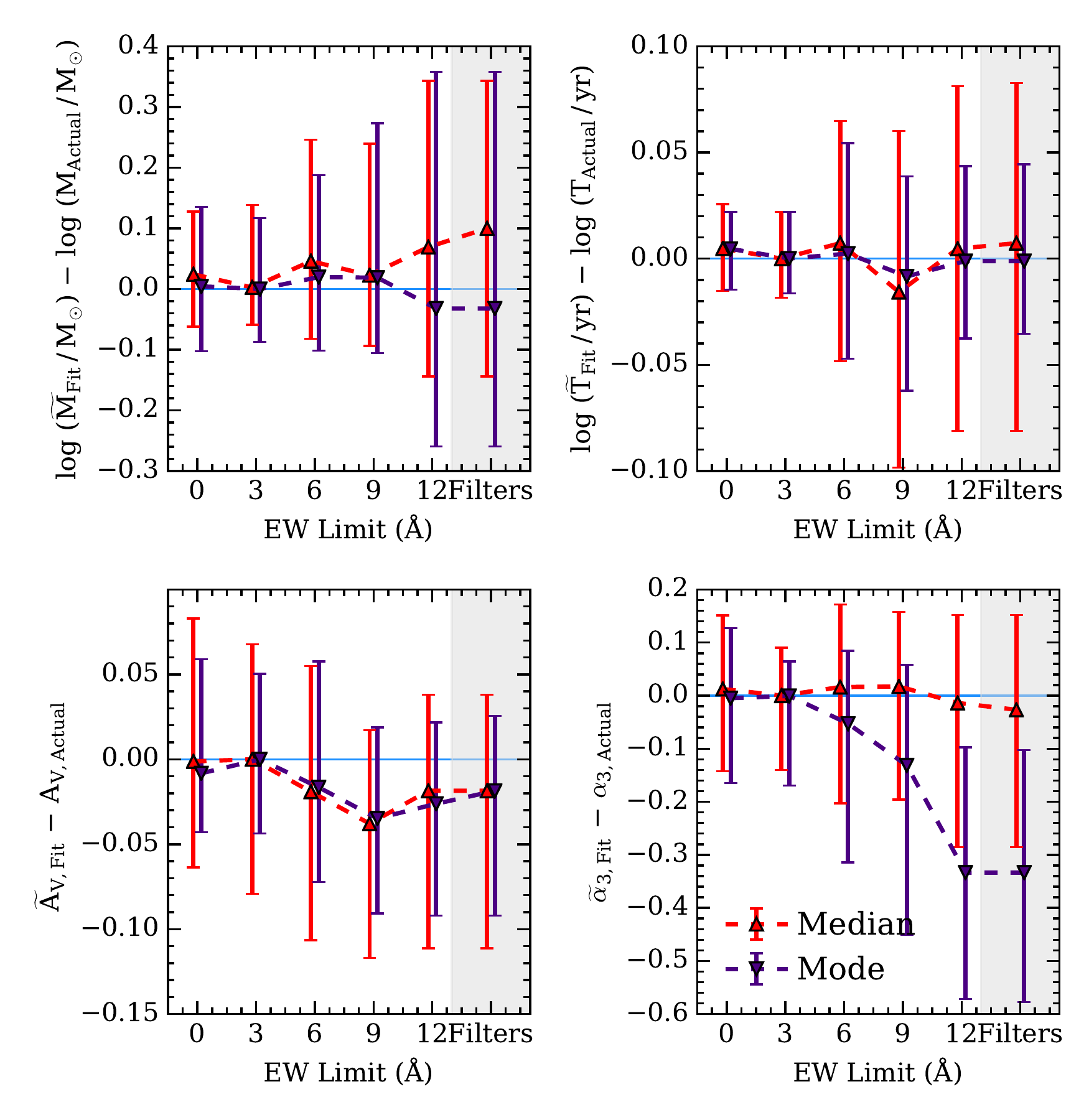} &
\includegraphics[scale=0.505,trim={0.43cm 0 0.4cm 0},clip]{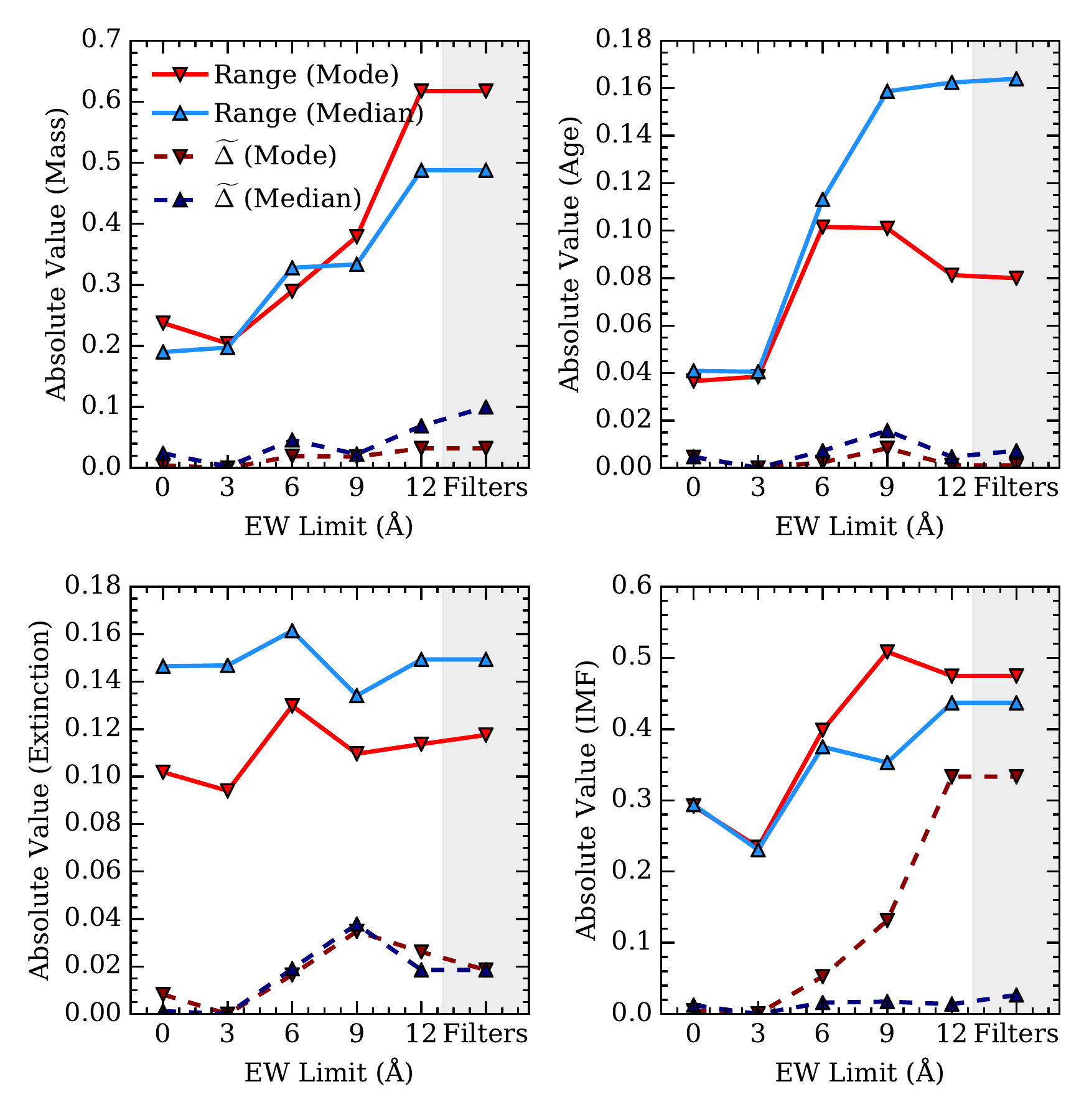} \\
(a) & (b)\\[6pt]

\end{tabular}
\end{center}
\caption{As Figure~\protect\ref{fig:ewlimsresid} but for the narrower age-range models detailed in Table~\protect\ref{tab:modelparams2}.}
\label{fig:ewlimsresidobs2}
\end{figure*}

In Figure~\ref{fig:ewlimsresidobs2} we repeat the sensitivity tests shown previously in Figure~\ref{fig:ewlimsresid}, this time using the models covering a narrower range of ages. The results are similar to what was seen in Figure~\ref{fig:ewlimsresid}, with similar improvements in mass and \imfs, and improved recovery of the cluster age. Compared to the interquartile range of the median residuals on \imfs\ shrinking by $32\%$, with the restricted range of ages we see the same level of improvement. This improves to $61\%$ when we include all the spectral information. 

\label{lastpage}

\end{document}

%% file: table_models.tex
\begin{table*}
\begin{center}
  \begin{tabular}{ l  c c c c c  c c c c}
    \hline
    Name  & $M_{*,{\rm Max}}$ & Z & $\log M $ & $ \log t $ & \#$^{\rm{a}}$  & \imfs \\ 
     & &  & ($\msun$) & (yr) & &  &  & \\ \hline \vspace{0.1cm}
     
Library   &  120  & 0.020 & 2.7--5.7\,(C) & 4.7--7.7\,(C) & $ 5 \times 10^7 $ & $-3.2$ -- $-1.3$\,(C)\\
Mock Observations &  120  & 0.020 & 3--5\,(C) & 5--7\,(C) & $ 5 \times 10^2  $ & $-3.0$ -- $-1.5$\,(C)\\
Set 1 : High Cutoff, Solar $Z$    &  120  & 0.020 & 3,6\,(D)           & 5,6,7\,(D)     & $ 10^3 $ & $-3.0,-2.3, -1.5$\,(D)\\
Set 2 : High Cutoff, High $Z$   &  120  & 0.050 & 3,6 \,(D)         & 5,6,7\,(D)     & $ 10^3 $ & $ -2.3$ \\ 
Set 3 : High Cutoff, Low $Z$    &  120  & 0.004 & 3,6\,(D)           & 5,6,7\,(D)     & $ 10^3 $ & $ -2.3$ \\ 
Set 4 : Low Cutoff, Solar     &  30   & 0.020 & 3,6\,(D)           & 5,6,7\,(D)     & $ 10^3 $ & $-3.0,-2.3, -1.5$\,(D)\\
Set 5 : Low Cutoff, High $Z$    &  30   & 0.050 & 3,6\,(D)           & 5,6,7\,(D)     & $ 10^3 $ & $ -2.3$ \\ 
Set 6 : Low Cutoff, Low $Z$    &  30   & 0.004 & 3,6\,(D)           & 5,6,7\,(D)     & $ 10^3 $ & $-2.3$ \\ 
Squares &  120  & 0.020 & 3,6\,(D)           & 5--7\,(C) & $ 10^5 $ & $-3.0$ -- $-1.5$\,(C)\\
    \hline
\multicolumn{9}{l}{ $^{\rm{a}}$ Number of realisations}
  \end{tabular}
\end{center}
\caption{Table of the parameters used in \slug\ for generating the mock star clusters used within this work. The base IMF used for all these model sets is the 3 component broken power law of the Kroupa IMF. The break points between the segments of this power law are situated at  0.01, 0.08, 0.5, $M_{*,{\rm Max}}$. $Z$ is the metallicity. The (D) represents discrete sampling, with \# realisations at each value, and (C) represents continuous sampling, with \# realisations drawn from a flat distribution across the given range.}
\label{tab:modelparams}
\end{table*}

%% file: table_lines.tex
\begin{table}
\begin{center}
  \begin{tabular}{ l | c | c | c }
    \hline
    Feature Name   & Integration Range & $\phi^{\rm a}$ \\ 
                   & $({\rm \AA})$       & $({\rm eV})$ &\\ 
    \hline \vspace{0.1cm}
    O{\sc vi}      & $1015.0-1050.0$         & $138.1189$ \\
    \qquad O{\sc vi}a$^{\dagger}$     & $1015.0-1037.0$         &  --           \\
    \qquad O{\sc vi}b     & $1037.0-1050.0$         &  --          \\
    C{\sc iii}$^{\dagger}$     & $1174.0-1177.0$         & $47.88778$ \\
    Si{\sc iii}b   & $1205.0-1207.0$         & $33.49300$ \\
    N{\sc v}       & $1220.0-1250.0$         & $97.89013$ \\
    \qquad N{\sc v}a$^{\dagger}$      & $1220.0-1240.0$         &    --     \\
    \qquad N{\sc v}b      & $1240.0-1250.0$         &    --       \\
    Bl1302 *        & $1292.0-1312.0$         &   --       \\
    C{\sc ii}$^{\dagger}$      & $1334.0-1338.0$         & $24.3845$  \\
    O{\sc v}       & $1355.0-1375.0$         & $113.8989$ \\
    \qquad O{\sc v}a$^{\dagger}$      & $1355.0-1370.0$         &    --        \\
    \qquad O{\sc v}b      & $1370.0-1375.0$         &    --        \\
    Si{\sc iv} * $^{\dagger}$    & $1387.0-1407.0$         & $45.14179$ \\
    Bl1425 *        & $1415.0-1435.0$         &   --       \\
    Si{\sc iii}a $^{\dagger}$  & $1416.0-1418.0$         & $33.49300$ \\
    Fe1453 * $^{\dagger}$       & $1440.0-1466.0$         &   --       \\
    C{\sc iv}      & $1527.0-1569.0$         & $64.49351$ \\
    \qquad C{\sc iv}a * $^{\dagger}$    & $1530.0-1550.0$         &   --       \\
    \qquad C{\sc iv}c *    & $1540.0-1560.0$         &   --       \\
    \qquad C{\sc iv}e *    & $1550.0-1570.0$         &   --       \\
    Bl1617 *        & $1604.0-1630.0$         &   --       \\
    He{\sc ii} $^{\dagger}$    & $1639.5-1641.0$         & $54.4177650$ \\
    Bl1664 *        & $1651.0-1677.0$         &   --       \\
    N{\sc iv} $^{\dagger}$     & $1703.0-1727.0$         & $77.4735$  \\
    Bl1719 *        & $1709.0-1729.0$         &   --       \\
    Bl1853 *        & $1838.0-1868.0$         &   --       \\
    Si{\sc i} $^{\dagger}$     & $1843.0-1846.0$         & $8.15168$  \\
    Fe{\sc ii}a *    & $2382.0-2422.0$         & $16.19920$ \\
    Bl2538 *        & $2520.0-2556.0$        &    --      \\
    Fe{\sc ii}b *   & $2596.0-2622.0$         & $16.19920$ \\
    MgWide *        & $2670.0-2870.0$         &   --       \\      
    Mg{\sc ii} * $^{\dagger}$   & $2784.0-2814.0$         & $15.035271$ \\
    Mg{\sc i} * $^{\dagger}$    & $2839.0-2865.0$        & $7.646236$ \\   
    \hline
\multicolumn{4}{l}{ $^{\rm{a}}$ Ionisation potential}
  \end{tabular}
\end{center}
\caption{Table of the line features used in this work, where we list the integration regions used for calculating the equivalent width of each feature, along with an ionisation energy if appropriate. The ionisation potentials given were obtained from \protect\cite{NIST_ASD}. The lines are sorted by increasing leading edge wavelength of their integration region. Spectral features taken from \protect\cite{Fanelli1992-UVLibrary} are marked with an asterisk (*). In C{\sc iv}, the letters ``a'',``c'',``e'' represent the absorption, centre, and emission respectively (as given in \protect\cite{Fanelli1992-UVLibrary}). For other P-Cygni features such as N{\sc v}, the ``a'' and ``b'' represent the absorption and emission components respectively. For other features such as Si{\sc iii}, the ``a'' and ``b'' represent two occurrences of the same feature, with ``a'' being the stronger in equivalent width. Features marked with a $\dagger$ mark are those used in the full-set analysis as they are sufficiently independent of one another. Sub-regions of larger spectral features are indented.}
\label{tab:lines}
\end{table}

%% file: table_stacks.tex
\begin{table*}
\begin{center}
  \begin{tabular}{ l  c  c  c  c}
    \hline
    Range         & $\Delta\log M$          & $\Delta\log t$          & $\Delta A_{\rm V}$       & $\Delta$\imfs  \\ 
    (Percentiles) & ($\msun$)               & (yr)                    &                          &                \\ 
    \hline \vspace{0.1cm}                             
    5-95 (MC)         & $-0.61$~($-35\%$)       & $-0.63$~($-49\%$)       & $+0.18$~($+31\%$)  & $-0.34$~($-23\%$)    \\ 
    25-75 (MC)        & $-0.30$~($-42\%$)       & $-0.18$~($-41\%$)       & $+0.08$~($+36\%$)  & $-0.24$~($-35\%$)    \\    
    5-95  (TC)       & $-0.84$~($-39\%$)       & $-0.87$~($-55\%$)       & $+0.14$~($+20\%$)  & $-0.50$~($-26\%$)    \\ 
    25-75 (TC)        & $-0.33$~($-41\%$)       & $-0.22$~($-45\%$)       & $+0.07$~($+27\%$)  & $-0.24$~($-32\%$)    \\
    \hline
  \end{tabular}
\end{center}
\caption{Table of the difference in inter-quartile and inter-percentile ranges for the median-centred (MC) and  true-centred stacks (TC) shown in Figure~\ref{fig:2dstacks}, changing from the case with photometry alone to the case including both photometry and the selected UV features. The numbers in parentheses correspond to the percentage change in the ranges, compared to the original value. }
\label{tab:ipr}
\end{table*}

%% file: table_models_2.tex
\begin{table*}
\begin{center}
  \begin{tabular}{ l    c c c  c c c c}
    \hline
    Name    & $\log M $ & $ \log t $ & \#$^{\rm{a}}$  & \imfs \\ 
     &   ($\msun$) & (yr) & &  &  & \\ \hline \vspace{0.1cm}
     
Library    & 2.7--5.7\,(C) & 5.7--7.7\,(C) & $ 5 \times 10^7 $ & $-3.2$ -- $-1.3$\,(C)\\
Mock Observations   & 3--5\,(C) & 6--7\,(C) & $ 5 \times 10^2  $ & $-3.0$ -- $-1.5$\,(C)\\
    \hline
\multicolumn{7}{l}{ $^{\rm{a}}$ Number of realisations}
  \end{tabular}
\end{center}
\caption{Table of the parameters used in \slug\ for generating the higher-density library and mock observations for the model clusters with a narrower range of ages. The base IMF used for all these model sets is the 3 component broken power law of the Kroupa IMF. The break points between the segments of this power law are situated at  0.01, 0.08, 0.5, $M_{*,{\rm Max}}$. $Z$ is the metallicity. The (D) represents discrete sampling, with \# realisations at each value, and (C) represents continuous sampling, with \# realisations drawn from a flat distribution across the given range. All simulations were run with solar metallicity, and an IMF cutoff of $M_{*,{\rm Max}}=120\msun$}
\label{tab:modelparams2}
\end{table*}

%% file: table_stacks2.tex
\begin{table*}
\begin{center}
  \begin{tabular}{ l  c  c  c  c}
    \hline
    Range         & $\Delta\log M$          & $\Delta\log t$          & $\Delta A_{\rm V}$       & $\Delta$\imfs  \\ 
    (Percentiles) & ($\msun$)               & (yr)                    &                          &                \\ 
    \hline \vspace{0.1cm}                             
    5-95 (MC)         & $-0.45$~($-26\%$)       & $-0.33$~($-38\%$)       & $+0.17$~($+26\%$)  & $-0.25$~($-16\%$)    \\ 
    25-75 (MC)        & $-0.22$~($-31\%$)       & $-0.08$~($-29\%$)       & $+0.06$~($+24\%$)  & $-0.20$~($-27\%$)    \\    
    5-95  (TC)       & $-0.63$~($-29\%$)       & $-0.56$~($-50\%$)       & $+0.15$~($+19\%$)  & $-0.36$~($-18\%$)    \\ 
    25-75 (TC)        & $-0.25$~($-31\%$)       & $-0.09$~($-29\%$)       & $+0.05$~($+17\%$)  & $-0.20$~($-25\%$)    \\
    \hline
  \end{tabular}
\end{center}
\caption{Table of the difference in inter-quartile and inter-percentile ranges for the median-centred (MC) and  true-centred stacks (TC) shown in Figure~\ref{fig:2dstacks2}, changing from the case with photometry alone to the case including both photometry and the selected UV features. The numbers in parentheses correspond to the percentage change in the ranges, compared to the original value. The models included in this analysis are limited to ages of $10^{6-7}\,\yr$, as shown in Table~\ref{tab:modelparams2}. }
\label{tab:ipr2}
\end{table*}